\documentclass[natbib]{emulateapj-rtx4}
\usepackage{graphicx}
\usepackage{hyperref}

\newcommand{\icarus}{Icarus}

\slugcomment{Technical Report for CTIO, July 2012.
Last updated 1 March 2013}
\shorttitle{Coordinates for Cerro Tololo and Cerro Pach\'{o}n}
\shortauthors{Mamajek}
\begin{document}

\title{Accurate Geodetic Coordinates for Observatories\\ on Cerro
  Tololo and Cerro Pach\'{o}n}

\author{Eric E. Mamajek\altaffilmark{1,2}} 
\altaffiltext{1}{Cerro Tololo Inter-American Observatory,
Casilla 603, La Serena, Chile and National Optical Astronomy Observatory,
950 N. Cherry Ave., Tucson, AZ, 85721}
\altaffiltext{2}{Current address: University of Rochester, Department of Physics \& 
Astronomy, Rochester, NY, 14627-0171, USA} 
\email{emamajek@pas.rochester.edu}

\begin{abstract} 
As the 50th anniversary of the Cerro Tololo Inter-American Observatory
(CTIO) draws near, the author was surprised to learn that the
published latitude and longitude for CTIO in the Astronomical Almanac
and {\it iraf} observatory database appears to differ from modern
GPS-measured geodetic positions by nearly a kilometer.  Surely, the
position for CTIO could not be in error after five decades? The source
of the discrepancy appears to be due to the $\sim$30'' difference
between the astronomical and geodetic positions -- a systematic effect
due to vertical deflection first reported by Harrington, Mintz Blanco,
\& Blanco (1972).  Since the astronomical position is not necessarily
the desired quantity for some calculations, and since the number of
facilities on Cerro Tololo and neighboring Cerro Pach\'{o}n has grown
considerably over the years, I decided to measure accurate geodetic
positions for all of the observatories and some select landmarks on
the two peaks using GPS and Google~Earth.  Both sets of measurements
were inter-compared, and externally compared to a high accuracy
geodetic position for a NASA Space Geodesy Program survey monument on
Tololo. I conclude that Google~Earth can currently be used to
determine absolute geodetic positions (i.e.  compared to GPS) accurate
to roughly $\pm$0.15'' ($\pm$5~m) in latitude and longitude without
correction, or approximately $\pm$0''.10 ($\pm$3~m) with correction.
I tabulate final geodetic and geocentric positions on the WGS~84
coordinate system for all astronomical observatories on Cerro Tololo
and Cerro Pach\'{o}n with accuracy $\pm$0''.1 ($\pm$3~m).  One
surprise is that an oft-cited position for LSST is in error by 9.4 km
and the quoted elevation is in error by 500 m.
\end{abstract}


\section{Motivation}

For 50 years, the Association of Universities for Research in
Astronomy
(AURA)\footnote{\href{http://www.aura-astronomy.org/}{http://www.aura-astronomy.org/}}
has supported a growing number of astronomical observatories on
\href{https://maps.google.com/maps?&z=16&q=-30.1690556,-70.8062861}{Cerro
  Tololo} and
\href{https://maps.google.com/maps?&z=16&q=-30.2407416667,-70.7366833}{Cerro
  Pach\'{o}n}. These observatories include ones operated by the
National Optical Astronomy Observatories
(NOAO)\footnote{\href{http://www.ctio.noao.edu/noao/}{http://www.ctio.noao.edu/noao/}
  and \href{http://www.noao.edu/}{http://www.noao.edu/}} and several
other ``tenant'' scientific research facilities.  Since the founding
of the Cerro Tololo Inter-American Observatory and purchase of the
``El Totoral'' property 25 November 1962\footnote{``Brief History of
  the Cerro Tololo Inter-American Observatory'' written by Victor
  Blanco in February 1993:
  \href{http://www.ctio.noao.edu/noao/content/ctio-history}{http://www.ctio.noao.edu/noao/content/ctio-history}},
CTIO has provided a critical platform for astronomical observations of
the southern skies\footnote{Including, of course, support for
  observations with the Blanco 4-m telescope for the High-Z Supernova
  Search team and Supernova Cosmology Project which led to the 2011
  Nobel Prize in physics
  \citep[e.g.][]{Riess98,Schmidt98,Perlmutter99}, and important
  pioneering work by the Cal\'{a}n/Tololo SNe Ia survey
  \citep[e.g.][]{Hamuy96}. CTIO continues to provide support for the
  investigation of the cosmological acceleration through the Dark
  Energy Survey (DES) and Dark Energy Camera (DECam) project on the
  Blanco 4-m telescope.}, and continues to support impressive science
on no less than two dozen facilities.  The neighboring peak Cerro
Pach\'{o}n has been developed over the past two decades, hosting the
SOAR\footnote{\href{http://www.soartelescope.org/}{http://www.soartelescope.org/}}
and Gemini
South\footnote{\href{http://www.gemini.edu/}{http://www.gemini.edu/}}
telescopes.  The Large Synoptic Survey Telescope
(LSST)\footnote{\href{http://www.lsst.org/lsst/}{http://www.lsst.org/lsst/}}
is currently under construction on the
\href{https://maps.google.com/maps?&z=18&q=-30.2446333333,-70.749416667}{El
  Pe\~{n}\'{o}n peak of Cerro Pach\'{o}n}.  With excellent seeing,
weather, monitoring and mitigation of light pollution, and strong
legacy of development of infrastructure and trained staff, the
\href{https://maps.google.com/maps?&z=16&q=-30.1690556,-70.8062861}{Cerro
  Tololo} and
\href{https://maps.google.com/maps?&z=16&q=-30.2407416667,-70.7366833}{Cerro
  Pach\'{o}n} sites are well positioned to host astronomical
observatories for the next half century.

The precise positions of the observatories on Cerro Tololo has been
historically useful for a variety of purposes, including lunar
occultations \citep[e.g.][]{Lasker73,Vilas77}, planetary occultations
\citep[e.g.][]{French83,Hubbard97}, planetary ring occultations
\citep[e.g.][]{Elliot81}, and asteroid astrometry
\citep[e.g.][]{Buie12}.  Among other purposes, precise positions for
telescopes on these peaks are important for applying velocity
corrections for precise radial velocity measurements with the CHIRON
spectrograph on the SMARTS 1.5-m telescope \citep[][]{Schwab10}, and
will be important for accounting for geocentric parallax when
determining orbits of nearby small solar system bodies imaged with the
Dark Energy Camera (DECam) on the Blanco 4-m telescope and LSST.

Given the importance of a precise position for astronomical
calculations, it came as a surprise to the author that the position
for CTIO published in the Astronomical Almanac and in {\it iraf}
obsdb.dat file (observatory database) differed from GPS-measured
geodetic positions on Cerro Tololo by approximately a kilometer. As I
discuss later, this is not an ``error'' per se. The discrepancy arises
due to the mismatch in definitions between astronomical and geodetic
coordinates \citep{Harrington72,Blanco72}. However, the issue raised
awareness that {\it (1) there are conflicting (and sometimes
  incorrect) coordinates and elevations for facilities on Cerro Tololo
  and Cerro Pach\'{o}n on the WWW and in the literature, and (2)
  coordinates have not yet been estimated for many of the new
  facilities on these peaks}. Hence, an accessible review of the
coordinates for the Tololo and Pach\'{o}n facilities was long overdue.

This document is a summary of my notes and measurements for the
coordinates of facilities on Cerro Tololo and Cerro Pach\'{o}n.  It
may be updated later if improved measurements or corrections come to
light. After some discussion on terrestrial coordinate systems (\S2),
historical review and document archaeology (\S3), I present new
measurements of accurate geodetic and geocentric coordinates for the
astronomical facilities on Cerro Tololo and Cerro Pach\'{o}n (\S4),
and intercompare the measurements and assess their accuracy (\S5). The
best estimates of the coordinates for the observatories are compiled
in Table \ref{tab:final}. The document was first written in mid-2012,
and a summary of revisions is provided after the bibliography.

\section{Background on Coordinate and Elevation Systems \label{background}}


Following the ISO 6709 standard, geodetic positions in this paper are
quoted as latitude ($\phi$), longitude ($\lambda$), and elevation
($H$) on the World Geodetic System 84 (WGS~84) coordinate frame,
unless otherwise noted. Elevations will be discussed later in this
section, as they require a more detailed explanation. North latitude
and east longitude are positive.  Longitude 0$^{\circ}$ on the WGS~84
frame corresponds to the International Reference Meridian defined by
the International Earth Rotation and Reference Systems Service (IERS),
which lies approximately 5'' ($\sim$100 m) east of the meridian at the
Royal Observatory in Greenwich.  One degree ($^{\circ}$) of latitude
$\phi$ is 110.86 km, one minute ($^{'}$) is 1.85 km, and one second
($^{''}$) is 30.8 m.  At the latitude of Cerro Tololo, one degree of
longitude $\lambda$ is 96.32 km, one minute ($^{'}$) is 1.61 km, one
second is 26.8 m.


Global Position System (GPS; also called NAVSTAR) is the well known
navigation system supported by a constellation of 24 orbiting
satellites launched by the U.S. Department of Defense
\citep[e.g.][]{GPS92}.  Altitudes measured by GPS are referenced to a
gravitational equipotential surface that defines sea level: the World
Geodetic System 84 (WGS~84).  According to a Department of Defense
publication tilted ``Military Standard for Department of Defense World
Geodetic System (WGS)''\footnote{Document MIL-STD-2401, dated 11~January~1994,
  \href{http://earth-info.nga.mil/publications/specs/printed/2401/2401.pdf}{http://earth-info.nga.mil/publications/specs/printed/2401/2401.pdf}},
the WGS~84 provides ``{\it the basic reference frame (coordinate
  system), geometric figure for the earth (ellipsoid), earth
  gravitational model, and means to relate positions on various
  geodetic datums and systems for DoD operations and applications.}''
The WGS~84 geometric figure is an oblate spheroid with radius at the
equator of 6378137 m, and a flattening of $f$ = 1/298.257223563 (with
a corresponding polar radius of approximately 6356752 m. While there
is a 1984 version of the geoid model, there have been later updates of
the Earth Gravitational Model (EGM; e.g. EGM96)\footnote{For further
  details on WGS~84 and its minor alterations over the past decades,
  see NIMA Technical Report TR8350.2, "Department of Defense World
  Geodetic System 1984, Its Definition and Relationships With Local
  Geodetic Systems", 3rd edition, Amendment 1 (3 Jan 2000):
  \href{http://earth-info.nga.mil/GandG/publications/tr8350.2/tr8350_2.html}{http://earth-info.nga.mil/GandG/publications/tr8350.2/tr8350$\_$2.html}.}
-- indeed improving the resolution and accuracy of the geoid model is
a never-ending pursuit of the geodesy
community\footnote{\href{http://www.iag-aig.org/}{http://www.iag-aig.org/}}.


Quoted latitudes are usually {\it geodetic} ($\phi$), which measure
the angle between a normal at a position on a spheroid, and the
equatorial plane. Sometimes {\it geocentric} latitudes ($\phi^{'}$)
are quoted, measuring the angle between the line between the surface
position and the Earth's center, and the equatorial plane.  Geodetic
and geocentric longitudes $\lambda$ are identical as they share the
same axis and reference meridian.  Geodetic latitudes ($\phi$) and
geocentric latitudes ($\phi^{'}$) at Tololo and Pach\'{o}n differ by
$\sim$10'. Conversions for geodetic and geocentric coordinates can be
found in Sec. 4 of the Explanatory Supplement to the Astronomical
Almanac \citep{Seidelmann92}.


One can define {\it geodetic height} with respect to the reference
ellipsoid.  Elevations are classically defined with respect to {\it
  mean sea level}, however this does not follow the ellipsoid exactly
due to the local concentration of mass -- and of course defining sea
level on land is not trivial. Earth gravitational models (EGMs) are
mathematical approximations for the {\it geoid}, the Earth's
gravitational equipotential surface.  The gravitational vector is
perpendicular to the geoid. The ocean's mean sea level roughly follows
the geoid, so one needs to define the geoid in order to quote
altitudes with respect to ``mean sea level'' on land, as is commonly
done. The geoid varies from the ellipsoid shape by deviations of up to
approximately $\pm$100 m.  In reality, the geoid is a very complicated
and irregular surface following local concentrations of mass.  In
practice, it is defined by high order spherical harmonic expansion
expressions.  One commonly used EGM is Earth Gravitational Model 1996
(EGM96).  This defines the geoid with a spherical harmonic series of
$n$ = 360 ($\sim$100 km resolution). There are three different
``heights'' of note: {\it geodetic height} ($h$), orthometric height
or {\it elevation} ($H$), and {\it geoid height} ($N$). They are
related as:

\begin{equation}
H = h - N
\end{equation}

i.e. the elevation $H$ is equal to the geodetic height $h$ minus the
geoid height $N$ \citep[measured with respect to the reference
ellipsoid; see Chapter 4 of][]{Seidelmann92}.

An online tool from
UNAVCO\footnote{\href{http://www.unavco.org/community_science/science-support/geoid/geoid.html}{http://www.unavco.org/community$\_$science/science-support/geoid/geoid.html}}
was used to estimate the geoid height $N$ on the EGM96 model for Cerro
Tololo and Cerro Pach\'{o}n. The mean geoid heights $N$ for Tololo and
Pach\'{o}n are approximately 34.6 and 35.0 meters, respectively.
Geoid heights at individual observatories will be tabulated at the end
of the paper.  EGM96 geoid heights across both peaks vary at the
$<$0.1 m level from site to site. There are subtle differences in
geoid models as they have improved in accuracy and resolution over the
years. For example, at the position of the Blanco 4-m telescope, the
(EGM84, EGM96, and
EGM2008\footnote{\href{http://earth-info.nga.mil/GandG/wgs84/gravitymod/egm2008/index.html}{http://earth-info.nga.mil/GandG/wgs84/gravitymod/egm2008/
    index.html}}\,\footnote{\href{http://geographiclib.sourceforge.net/cgi-bin/GeoidEval}{http://geographiclib.sourceforge.net/cgi-bin/GeoidEval}})
Earth gravitational models predict geoid heights $N$ = (31.41, 34.62,
33.31) meters.  Hence, the definition of ``mean sea level'' under the
observatories has varied at the $\pm$few meter level over the past
decades.  Throughout this paper, all geoid heights $N$ refer to the
EGM96 model, unless otherwise noted.


{\it To what degree are Cerro Tololo and Cerro Pach\'{o}n moving due
  to plate tectonics?} While the common perception is that Chile is
moving westward with the South American tectonic plate, there are
surprisingly large variations in the plate motion in Chile. These
large variations are traced through time series GPS measurements of
ground stations in Chile maintained by the Jet Propulsion
Laboratory\footnote{\href{http://sideshow.jpl.nasa.gov/mbh/series.html}{http://sideshow.jpl.nasa.gov/mbh/series.html}}.
Generally, stations north of Santiago (Santiago, Valpara\'{i}so,
Copiap\'{o}, Iqueque) are moving {\it northeast} with respect to
WGS~84 at $\sim$30 mm/yr, while Chilean stations south of Santiago are
moving {\it northwest} at $\sim$90 mm/yr.  Although I have been unable
to find accurate geodetic motions for Cerro Tololo and Cerro
Pach\'{o}n in particular, judging by the measurements for other
well-studied benchmarks in Chile, it is very likely that the peaks are
moving at $<$100 mm/yr, and most likely $<$30 mm/yr.
UNAVCO\footnote{\href{http://www.unavco.org/community$\_$science/science-support/crustal$\_$motion/dxdt/model.html}{http://www.unavco.org/community$\_$science/science-support/crustal$\_$motion/dxdt/model.html}}
provides a plate motion calculator (model GSRM v1.2) on their website,
and at Cerro Tololo's position, they predict that the South American
plate is moving with respect to the WGS~84 coordinate system at 9.24
mm/yr (9.17 mm/yr N, 1.10 mm/yr E) - however this velocity is
substantially smaller than those reported for northern Chile JPL
stations. Hence, over the five decade history of the observatory, the
site has very likely moved $<$5 m (and indeed the UNAVCO plate motion
calculator would predict only $\sim$0.5 m of motion over 50 years).
In summary, any deviations in the published positions for the
observatory larger than these amounts are unlikely to be attributable
to tectonic motion, especially on short timescales.

\section{Previous Coordinates and Elevations}

\subsection{Coordinates} 

Past published positions for Cerro Tololo and Cerro Pach\'{o}n and
individual structures are listed in Table \ref{tab:old}. The
Astronomical Almanac \citep[AA; e.g.][]{AA2013} often listed positions
for individual telescopes at major observatories during the early
1980's, but only listed positions for 4 telescopes on Tololo in the
1982 edition: the Blanco 4-m reflector, the 1.5-m reflector, the 1-m
reflector, and the 24'' Curtis Schmidt telescope. These positions were
presumably adopted from the CTIO facilities manual
\citep[e.g.][]{Walker80}.  The AA ceased listing positions for
individual telescopes at observatories after the mid-80's, and
provided only mean positions for observatories thereafter. Note that
the AA concedes that the observatory positions are a mix of geodetic
and astronomical values, and the type of position is not provided.

\begin{deluxetable*}{llllllllll}
\tabletypesize{\scriptsize} 
\tablecaption{Published Coordinates for
  Cerro Tololo and Cerro Pach\'{o}n\label{tab:old}}
\tablewidth{0pt} 
\tablehead{ \colhead{Site} & \colhead{Ref.} & \multicolumn{3}{c}{$\phi$} & \multicolumn{3}{c}{$\lambda$} & \colhead{Elev.} & \colhead{Notes}\\ 
\colhead{} & \colhead{} & \colhead{$\circ$} & \colhead{'} & \colhead{''} & \colhead{$\circ$} &
  \colhead{'} & \colhead{''} & \colhead{(m)} & \colhead{}} 
\startdata
Cerro Tololo & & & & & & & & & \\ 
\hline 
\href{https://maps.google.com/maps?&z=18&q=-30.166666667,-70.816666667}{CTIO} &\citet{Mayall68} & -30&10 &00 &-70&49 &00 & 2200 & $a$\\ 
\href{https://maps.google.com/maps?&z=18&q=-30.164722222,-70.815000000}{CTIO} &\citet{Blanco72} & -30&09 &53$\pm$1.8 &-70&48 &54$\pm$3& ...  & Astronomical, $b$\\ 
\href{https://maps.google.com/maps?&z=18&q=-30.165555555,-70.804444444}{CTIO} &\citet{Harrington72}&-30&09&56 &-70&48 &16 & 2210 & Topographic Map,$c$\\ 
\href{https://maps.google.com/maps?&z=18&q=-30.168944444,-70.805861111}{CTIO} &\citet{Walker80} & -30&10 &08.2$\pm$0.4 &-70&48&21.1$\pm$0.4& 2210 & Geodetic, $d$\\ 
\href{https://maps.google.com/maps?&z=18&q=-30.168944444,-70.806666667}{CTIO} &AA 1975-1980 & -30&10&08.2 &-70&48 &24 & 2399 &\\ 
\href{https://maps.google.com/maps?&z=18&q=-30.165277778,-70.815000000}{CTIO} &iraf & -30&09 &55 &-70&48 &54 & 2215 & $e$\\ 
\href{https://maps.google.com/maps?&z=18&q=-30.165000000,-70.815000000}{CTIO} &AA 1983-2013 & -30&09 &54 &-70&48 &54 & 2215 &\\ 
CTIO &MPC(2012) & ...&...&...  &-70&48 &21 & ...  &\\ 
\href{https://maps.google.com/maps?&z=18&q=-30.166055556,-70.814888889}{Blanco 4-m} &\citet{Walker80} & -30&09 &57.8 &-70&48 &53.6 & 2210 &\\ 
\href{https://maps.google.com/maps?&z=18&q=-30.166055556,-70.814888889}{Blanco 4-m} &AA 1982 & -30&09 &57.8 &-70&48 &53.6 & 2235 &\\ 
\href{https://maps.google.com/maps?&z=18&q=-30.166055556,-70.814888889}{Blanco 4-m} &\citet{Hubbard97}& -30&09 &57.8 &-70&48 &53.6 & 2235 &\\ 
\href{https://maps.google.com/maps?&z=18&q=-30.165638888,-70.815138889}{SMARTS 1.5-m} &AA 1982 & -30&09 &56.3 &-70&48 &54.5 & 2225 &\\ 
\href{https://maps.google.com/maps?&z=18&q=-30.165638888,-70.815138889}{SMARTS 1.5-m} &\citet{Walker80} & -30&09 &56.3 &-70&48 &54.5 & 2210 & Astronomical\\ 
SMARTS 1.3-m &MPC(2012) & ...&...&...  &-70&48 &21 & ...  & $f$\\ 
\href{https://maps.google.com/maps?&z=18&q=-30.16522222,-70.814416667}{SMARTS 1.0-m} &\citet{Walker80} & -30&09 &54.8 &-70&48 &51.9 & 2210 & Astronomical\\ 
\href{https://maps.google.com/maps?&z=18&q=-30.16522222,-70.814972222}{SMARTS 0.9-m} &\citet{Walker80} & -30&09 &54.8 &-70&48 &53.9 & 2210 & Astronomical\\ 
\href{https://maps.google.com/maps?&z=18&q=-30.1689444444,-70.80586111}{Schmidt 0.6-m}&\citet{Harrington72} & -30&10 &08.2$\pm$0.4&-70&48 &21.1$\pm$0.4 & 2399$\pm$10 & {\bf Geodetic}, $g$\\ 
\href{https://maps.google.com/maps?&z=18&q=-30.16541666667,-70.814638889}{Schmidt 0.6-m}&\citet{Harrington72} & -30&09 &55.5$\pm$1.4&-70&48 &52.7$\pm$2.0 & 2210 & {\bf Astronomical}, $g$\\ 
\href{https://maps.google.com/maps?&z=18&q=-30.16704027778,-70.816224444}{Tololo ``O''} &AURA survey notes& -30&10 &01.345 &-70&48 &58.408 & 2211.60 & $h$\\ 
\href{https://maps.google.com/maps?&z=18&q=-30.1724630333,-70.80004267778}{NASA Monument 7401}& NASA& -30&10 &20.86692 &-70&48 &00.15364& 2123.090 & Geodetic,$i$\\ 
\href{https://maps.google.com/maps?&z=18&q=-30.167638889,-70.80538889}{PROMPT} &PROMPT website & -30&10 &03.50 &-70&48 &19.40 & ... &\\ 
\href{https://maps.google.com/maps?&z=18&q=-30.16900000,-70.8040000}{LCOGT} &LCOGT website & -30&10 &08.4 &-70&48 &14.4 & 2200 & $j$\\ 
\hline 
Cerro Pach\'{o}n & & & & & & & & &\\ 
\hline 
\href{https://maps.google.com/maps?&z=18&q=-30.22833333,-70.72333333333}{Gemini S.} &AA 2013 & -30&13 &42 &-70&43 &24 & 2725 &\\ 
\href{https://maps.google.com/maps?&z=18&q=-30.22833333,-70.7233333333}{Gemini S.}  &iraf & -30&13&42 &-70&43 &24 & 2737 &\\ 
\href{https://maps.google.com/maps?&z=18&q=-30.24075000,-70.7366933333}{Gemini S.} &Gemini website & -30&14 &26.700&-70&44 &12.096 & 2722 & $k$\\ 
\href{https://maps.google.com/maps?&z=18&q=-30.23333333,-70.7166666667}{Gemini S.}  &\citet{Zombeck07}& -30&14 &...  &-70&43 &...  & 2715 &\\ 
\href{https://maps.google.com/maps?&z=18&q=-30.17225000,-70.800027778}{LSST} &\citet{Ivezic08B,Ivezic08} & -30&10 &20.1 &-70&48 &00.1& 2123 &\\ 
LSST &LSST website & ...&...&... &...&...&...  & 2647 & $l$\\ 
LSST Auxiliary  &LSST website & ...&...&... &...&...&...  & 2647 & $l$\\ 
\href{https://maps.google.com/maps?&z=18&q=-30.23800,-70.73372222}{SOAR} &\citet{Simms05} & -30&14 &16.8 &-70&44 &01.4 & 2738 &\\ 
\href{https://maps.google.com/maps?&z=18&q=-30.35000,-70.81666667}{SOAR} &\citet{Zombeck07}& -30&21 &...  &-70&49 &...  & 2701 &\\ 
SOAR &SOAR website & ...&...&...  &...&...&...  & 2701 & $m$\\ 
\href{https://maps.google.com/maps?&z=18&q=-30.2370536111,-70.734554444}{Pach\'{o}n monument (near SOAR)}& AURA survey notes&-30&14&13.393&-70&44&04.396&2724.60
\enddata 

\tablecomments{I list longitude as measured with east being positive,
  following the 2013 Astronomical Almanac \citep{AA2013}.  In the
  first column, hyperlinks are provided to plot the coordinates 
  on Google Maps for comparison. Notes: ($a$) \citet{Mayall68}
  position is quoted longitude as 4$^h$ 43$^m$ 16$^s$ W.  ($b$)
  \citet{Blanco72} geodetic position determined from observations of
  GEOS B satellite with Curtis Schmidt telescope from
  \citet{Harrington72} study.  \citet{Blanco72} also lists the
  \citet{Harrington72} astronomical coordinates.  ($c$)
  \citet{Harrington72} ``{\it approximate coordinates, taken from a
    topographic sheet based on the 1924 International Reference
    Ellipsoid and the 1956 Provisional South American Datum...}''.
  ($d$) \citet{Walker80} adopts the geodetic coordinates for the
  Schmidt telescope from \citet{Harrington72}.  ($e$) {\it iraf}
  position in obsdb.dat files: see e.g.
  \href{http://tdc-www.harvard.edu/iraf/rvsao/bcvcorr/obsdb.htm}{http://tdc-www.harvard.edu/iraf/rvsao/bcvcorr/obsdb.html}.
  ($f$) MPC longitude was listed as 289$^{\circ}$.1941, with
  $\rho$cos($\phi$) = 0.86560 and $\rho$sin($\phi$) = -0.49980 (see
  \href{http://www.minorplanetcenter.net/iau/lists/ObsCodes.html}{http://www.minorplanetcenter.net/iau/lists/ObsCodes.html}).
  ($g$) \citet{Harrington72} measured the astronomical position
  through timing the transits of stars with SAO catalog
  astrometry. They determined the geodetic position through measuring
  and timing positions of the GOES-II satellite (which had a
  well-determined orbit, and strobe lights which flashed sequences of
  7$\times$ 1.4 ms pulses every 4 s). The geodetic height is ``above
  ellipsoid'', (however {\it which} ellipsoid is not explicitly
  mentioned (but given that the observations analyzed in 1971, it is
  likely WGS 66).  ($h$) ``O'' is Tololo Control survey monument SE of
  Blanco 4-m.  Values from AURA survey notes from Don Cassidy which
  triangulated the position from Peralillo and Pach\'{o}n survey
  markers using theodolite observations, tied to elevations determined
  by Instituto Geographic Militar (IGM).  ($i$) NASA website (Space
  Geodesy Program):
  \href{http://cddis.nasa.gov/site$\_$cat/cerr.html}{http://cddis.nasa.gov/site$\_$cat/cerr.html}.
  ($j$) LCOGT website:
  \href{http://lcogt.net/site/cerro-tololo}{http://lcogt.net/site/cerro-tololo}.
  ($k$) Gemini website:
  \href{http://www.gemini.edu/sciops/telescopes-and-sites/locations}{http://www.gemini.edu/sciops/telescopes-and-sites/locations}.
  ($l$) LSST website:
  \href{http://www.lsst.org/lsst/science/summit_facilities}{http://www.lsst.org/lsst/science/summit$\_$facilities}.
  ($m$) SOAR website:
  \href{http://www.soartelescope.org/about-soar/location-1}{http://www.soartelescope.org/about-soar/location-1.}}
\end{deluxetable*}

The Minor Planet Center (MPC)
lists\footnote{\href{http://www.minorplanetcenter.net/iau/lists/ObsCodesF.html}{http://www.minorplanetcenter.net/iau/lists/ObsCodesF.html}}
two observatory code entries for CTIO: \#807 (Cerro Tololo
Observatory, La Serena) and \#I02 (Cerro Tololo, La Serena--2MASS).
At the time of writing (16 July 2012), the MPC lists identical
longitudes $\lambda$, and parallax constants $\rho$cos$\phi$ and
$\rho$sin$\phi$ for these two codes: $\lambda$ = 289$^{\circ}$.1941
(degrees east of Greenwich), and $\rho$cos$\phi$ = 0.86560 and
$\rho$sin$\phi$ = -0.49980. ``$\phi$'' in this case is apparently
geocentric latitude, and should be labeled $\phi^{'}$ to distinguish
it from geodetic latitude $\phi$.  From these last two quantities we
can calculate the parallax parameter $\rho$ to be 0.999532, and
latitude $\phi^{'}$ = -30$^{\circ}$.00228 = -30$^{\circ}$00'08''. This
is in good agreement with values derived later in this paper.

\subsection{1972 Harrington et al. Determination of Astronomical and
  Geodetic Coordinates \label{1972}}

In an annual report, \citet{Blanco72} reported the following regarding
the coordinates for CTIO:\\

{\bf ``{\it A program to determine the precise geodetic and geographic
    positions of Cerro Tololo was initiated by Dr. R. Harrington, U.S.
    Naval Observatory, Washington D.C., and Dr. and Mrs. V. M. Blanco,
    CTIO. Observations of the GEOS B satellite, which was flashed
    especially for this purpose, were made with the Curtis Schmidt
    telescope. The final reductions made at the Goddard Space Flight
    Center yield the following geodetic position:
    \href{https://maps.google.com/maps?&z=18&q=-30.16894444,-70.805861111}{$\phi$
      = -30$^{\circ}$ 10' 8''.2, $\lambda$ = W\,70$^{\circ}$ 48'
      21''.1}.  The geographic or astronomical position was determined
    by the method of equal altitudes with data obtained from a series
    of theodolite observations. The preliminary results are
    \href{https://maps.google.com/maps?&z=18&q=-30.164722222,-70.8150000}{$\phi$
      = -30$^{\circ}$ 9' 53''\,$\pm$\,1''.8, $\lambda$ =
      W\,70$^{\circ}$ 48' 54''\,$\pm$\,3''}. These figures suggest a
    deflection of the vertical of approximately 39 arc sec in the
    west-northwesterly direction, approximately perpendicular to the
    orientation of the Andean Cordillera and the deep off-shore
    Chile-Peru oceanic trench, which are the probable sources of the
    deflection.''}}\\

Examination of the original \citet{Harrington72} study\footnote{ A
  copy of the \citet{Harrington72} study proved exceedingly difficult
  to find, but the NOAO North library has a copy. A scanned copy can
  be downloaded at
  \href{http://www.pas.rochester.edu/$\sim$emamajek/Harrington72.pdf}{http://www.pas.rochester.edu/$\sim$emamajek/Harrington72.pdf}.}
suggests a typo in the deflection of vertical measured. The
astronomical and geodetic coordinates for the Curtis Schmidt telescope
on Tololo are listed in Table \ref{tab:old}. \citet{Harrington72}
concluded:\\

{\bf {\it ``The astronomic coordinates of the Curtis Schmidt telescope
    can be compared to the above geodetic coordinates to obtain the
    deflection of the vertical. This deflection amounts to
    30''.1\,$\pm$\,1''.7 towards an azimuth of
    295$^{\circ}$\,$\pm$\,3$^{\circ}$, which corresponds to
    approximately 930 meters on the ellipsoid surface.''}}\\

Hence, the source of the discrepancy between the Almanac and {\it
  iraf} positions for CTIO and what one would measure with a GPS or on
Google Earth appears to be due to the difference in the {\it type} of
latitude and longitude being reported (i.e. astronomical vs.
geodetic). While the astronomical position is useful for calculating
transit times, it may not be the position desired for other
calculations (e.g. taking into account geocentric parallax,
occultation calculations, etc.).

\subsection{1973 Survey Elevations \label{1973survey}}

I list in Table \ref{tab:elev} elevations for the concrete platforms
for observatories on the Tololo plateau from a 1973 survey map from
the NOAO Engineering Department\footnote{Contains note ``Survey map
  drawn by E.W. Ross, checked \& approved by Don Cassidy, Oct 12,
  1971.''}. The concrete platforms for all of the observatories on the
Tololo plateau have elevations ranging from 2209.60 m (SMARTS/Yale
1.0-m) to 2210.50 m (Blanco 4-m), i.e. less than $\Delta$$H$ = 1.1 m
among them. The elevations quoted in the survey are tied to topography
from the ``Fuerza Aerea de Chile - Servicio Aeoro Fotogrametrico
(November 1964)'' and ``supplemented by plane table topography by AURA
October 1966.''  The topography of the Tololo plateau was tied to
elevations of neighboring peaks with coordinates provided by the IGM
(including Cerro Pach\'{o}n and Cerro Peralillo) via a primary survey
monument called "Tololo", "Tololo Control", or "O" on the old survey
maps (discussed further in \S\ref{Tololo_Monument}). The Tololo survey
monument was assigned elevation 2211.60 m above sea level. Separate
notes from this period list an elevation of 2724.60 m for an IGM
survey monument on Cerro Pach\'{o}n, however it is not clear whether
this monument corresponds to one of the modern day survey monuments on
Pach\'{o}n as the site has been developed considerably in the
intervening four decades.

\begin{deluxetable*}{llr}
\tabletypesize{\scriptsize}
\tablecaption{Elevations for Observatory Platforms on 1973 Survey Map\label{tab:elev}}
\tablewidth{0pt}
\tablehead{
\colhead{Observatory} & \colhead{$H$} & \colhead{$\Delta$$H$}\\
\colhead{}            & \colhead{(m)} & \colhead{(m)}     
}
\startdata
Blanco 4-m (158'')            & 2210.50 & -1.10\\
SMARTS 1.5-m (60'')           & 2210.99 & -0.61\\
SMARTS 1.0-m (40'')           & 2209.60 & -2.00\\
SMARTS 0.9-m (36'')           & 2210.49 & -1.11\\
Curtis Schmidt 0.6-m (24'')   & 2210.00 & -1.60\\
No. 1 16'' (later USNO)       & 2210.31 & -1.29\\
No. 2 16'' (later CHASE)      & 2209.65 & -1.95\\
\hline
Tololo Control survey monument & 2211.60 & 0.00
\enddata
\tablecomments{$H$ is elevation scale measured with respect to sea
  level, tied to topography established by Fuerza Aerea de Chile -
  Servicio Aero Fotogrametrico, and supplemented by plane table
  topography measured by AURA. Later calculations show that the 1973
  survey elevations appear to assume a geoid within a meter or so of
  the EGM-84 model. $\Delta$$H$ is differential elevation with respect
  to Tololo Control survey monument. The mean elevation of the 7
  platforms is 2210.22 m. Their mean height on Google~Earth is 2202.7
  m. Their mean height measured via GPS is 2217.6 m. }
\end{deluxetable*}

\subsection{2008 Survey Elevations \label{2008survey}}

\begin{figure}
\epsscale{1.2}
\plotone{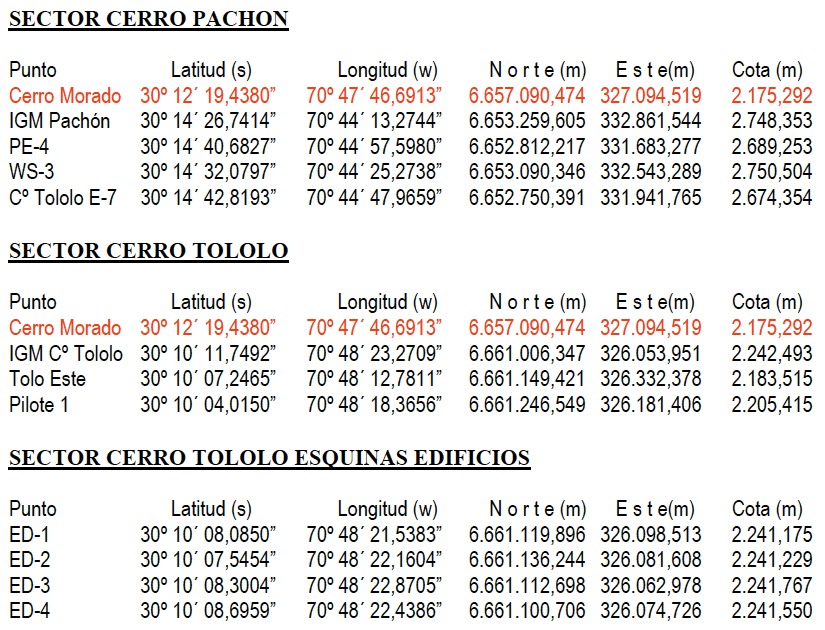}
\caption{Geodetic positions for survey monuments on Cerro Tololo, Pach\'{o}n, and Morado,
from a report completed by Juan Carlos Aravena Godoy for AURA dated 28 October 2008.  
The coordinates are on the WGS~84 system (as adopted by SIRAS, the Sistema de Referencia 
Geoc\'{e}ntrico para las Am\'{e}ricas), and elevations are with geodetic heights $h$ 
respect to the GRS~80
ellipsoid (i.e. not ``mean sea level'', since no geoid has been subtracted). Measurements
were taken via theodolite observations anchored to the Cerro Morado geodetic position (epoch 2002.0)
provided by the Wilfredo Rubio Dias, Depto.
  de Datos y C\'{a}lculo, Instituto Geographico Militar, Ejerctio de
  Chile (dated 23 Oct 2008). Pins ED-1 through ED-4 appear to correspond to positions
near the Schmidt telescope on Tololo. 
\label{fig:2008survey}}
\end{figure}

In Figure \ref{fig:2008survey}, I show a scanned table from a 2008
survey of Cerro Tololo, Cerro Pach\'{o}n, and the neighboring peak
\href{https://maps.google.com/maps?&z=18&q=-30.20539944444,-70.7963013138889}{Cerro
  Morado} (S of Tololo, NW of Pach\'{o}n), completed by Juan Carlos
Aravena Godoy for AURA. Cerro Morado has a Instituto Geographico
Militar (IGM) survey monument with well determined geodetic
coordinates, which were provided to the surveyers by the Chilean Army
(highlighted in red in Fig. \ref{fig:2008survey}).  The coordinates
are on the WGS~84 system (as adopted by SIRAS, the Sistema de
Referencia Geoc\'{e}ntrico para las Am\'{e}ricas), and elevations are
with respect to the GRS~80 ellipsoid.  The difference in geocentric
radii between the GRS~80 and WGS~84 ellipsoids at the position of the
Blanco 4-m is 26 $\mu$m (i.e.  10$^{-4.6}$ m). Hence, for all
practical purposes, the differences between the GRS~80 ellipsoid
adopted by SIRGAS (used in Chilean surveying) and the WGS~84
ellipsoids, are completely negligible
\footnote{The only difference between the WGS~84 and GRS~80 reference
  ellipsoid is due to the adopted flattening parameter ($f$) - in the
  {\it 6th} decimal place. So throughout, it is safe to assume that
  GRS~80 is equivalent to WGS~84. See also: NIMA Technical Report
  TR8350.2, "Department of Defense World Geodetic System 1984, Its
  Definition and Relationships With Local Geodetic Systems", 3rd
  edition, Amendment 1 (3 Jan 2000):
  \href{http://earth-info.nga.mil/GandG/publications/tr8350.2/tr8350_2.html}{http://earth-info.nga.mil/GandG/publications/tr8350.2/tr8350$\_$2.html}.}.
The 2008 survey also measured positions for 4 survey monuments near
the
\href{https://maps.google.com/maps?&z=18&q=-30.1690556,-70.8062861}{Schmidt}
telescope on the Tololo plateau. The elevations ranged from 2241.175
to 2241.767 m, i.e. $\Delta H$ = 0.652\,m, and $\bar{H}$ = 2241.430\,m
(elevations are with respect to the GRS~80 ellipsoid).  This is
probably representative of the GRS~80 heights of the concrete
platforms for the observatories on the Tololo plateau. The 1970s-era
survey notes also list elevations for 4 pins distributed around the
\href{https://maps.google.com/maps?&z=18&q=-30.1690556,-70.8062861}{Schmidt}
telescope, with elevations ranging from 2209.333 to 2209.985 m
($\Delta H$ = 0.592\,m; $\bar{H}$ = 2209.650\,m), and given the
description and close agreement in spread of elevations, these are
likely to be the same pins as measured in the 2008 survey.  Both the
Tololo Control and Schmidt survey monument measurements indicate that
the 1973 survey maps can be converted to be geodetic heights with
respect to the GRS~80 ($\approx$WGS~84) ellipsoid by adding
$\sim$31\,m. This difference is not far from the geoid undulations
predicted for Tololo's position in recent EGM models (\S 5.2.2).
These numbers seem to confirm that the 2008 survey elevations have not
subtracted off geoid undulation, and they are simply with respect to
the GRS~80 ellipsoid (hence, they are likely to have non-zero
elevations at mean sea level).

\section{Measurements}

\begin{figure}
\epsscale{1.0}
\plotone{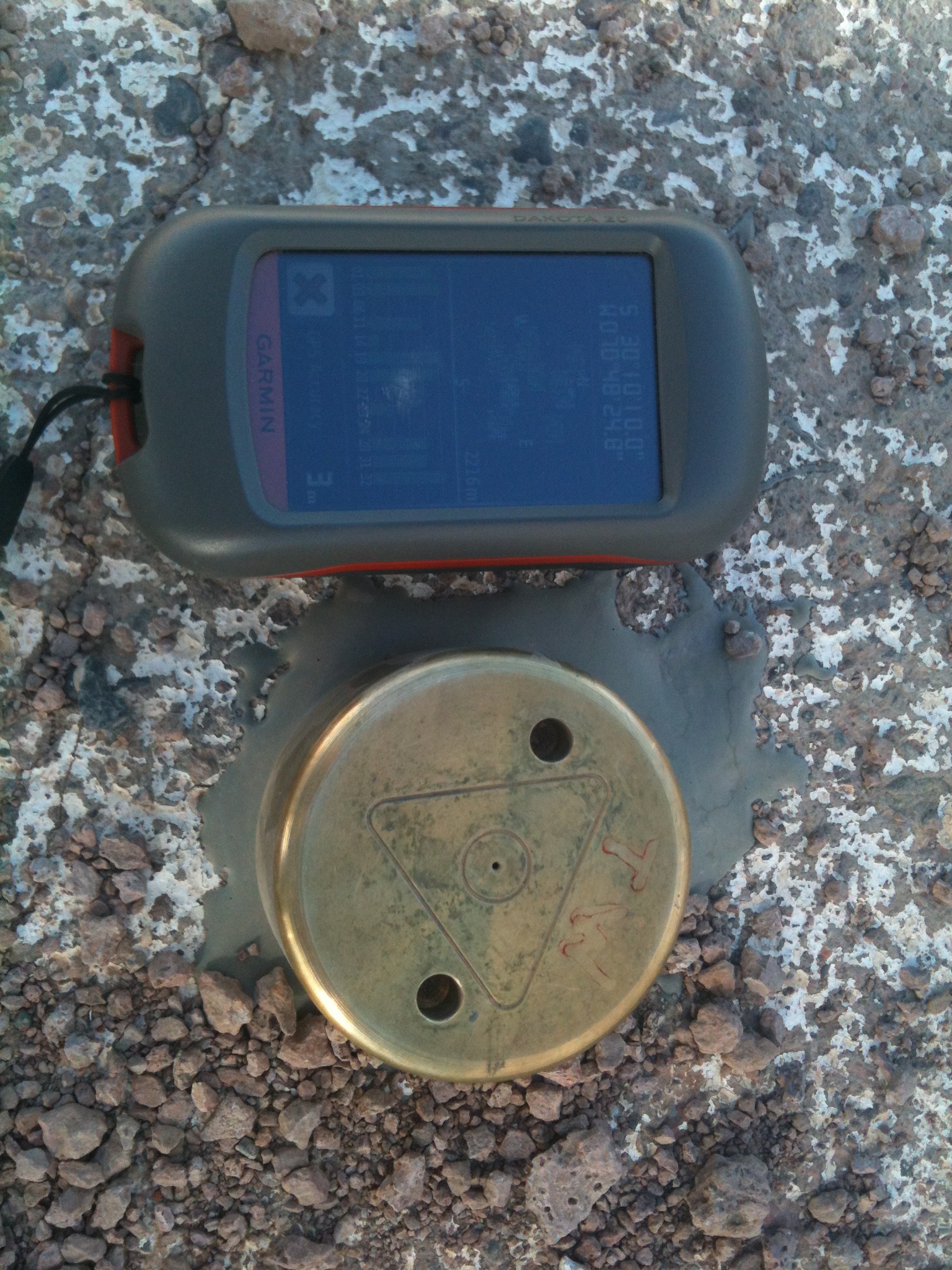}
\caption{The Garmin Dakota 20 GPS unit used in this study, sitting next
to the benchmark adjacent to the southwest corner of the concrete platform
for the SMARTS 1.5-m telescope. 
\label{GPS}}
\end{figure}

{\it GPS (Garmin):} Positions were measured with a Garmin brand Dakota
20 GPS\footnote{Courtesy of Michael Warner (CTIO).}, shown in Figure
\ref{GPS}. This GPS was set to update at 1 second intervals, quoted
positions to 0''.1 precision and elevations to 1 m precision, and
provided a regularly updated estimate of the position accuracy.
Typical quoted accuracy in a 1 sec interval ranged from as good as 3 m
(with unobstructed view and several satellites acquired), and at times
as bad as 20 m, close to the
advertised\footnote{\href{http://www8.garmin.com/aboutGPS/}{http://www8.garmin.com/aboutGPS/}}
mean accuracy of $\sim$15 m.

Measurements were taken in different modes. As the GPS would not
operate inside of domes due to obstructed view, I typically took two
sequences of measurements at four corners of a given structure, and
averaged the results. For assymetric structures (e.g. Gemini, SOAR), I
did some simple interpolation to estimate the central position.  The
GPS positions were very repeatable at the $\pm$0.1'' level (only very
rarely would subsequent measurements taken minutes apart differ by
0.2'').  Measurements were either taken (1) with the device sitting on
the flat cement platforms outside of the domes, or the ground if there
was no platform, or (2) with the device held in the hand,
approximately 1 meter above the ground. None of the elevation
measurements taken in this manner are quoted to better than 1 m
precision. For GPS measurements of geodetic benchmarks and observatory
sites currently lacking enclosures (e.g. KASI and T80-S), the GPS
device was left sitting on the ground at the position, and coordinates
were saved at 1 s or 5 s intervals for extended periods of time to the
GPS's memory.  As the GPS would often read out precisely the same
coordinates for several seconds on end (then jump small amounts,
presumably due to acquiring and losing individual satellites), it was
decided to measure statistical moments only on {\it unique} coordinate
sets in the time series data. Especially long time series measurements
were made of the benchmarks next to the SMARTS 1.5-m and NASA geodesy
benchmark next to the SARA (former Lowell 24'') telescope.

{\it GPS (iPhone):} Auxiliary GPS measurements were made with an
iPhone 3s using Entel phone service, and using the application {\it
  Compass}. The iPhone measurements suffered two major limitations:
sparse network coverage on the mountains, and the latitudes and
longitudes were listed at 1'' precision.  Elevation estimates were
also made using the iPhone application {\it Current
  Elevation}\footnote{\href{http://itunes.apple.com/us/app/current-elevation/id429847648}{http://itunes.apple.com/us/app/current-elevation/id429847648}}.
As the quoted positions were at lower precision than those provided by
the Garmin GPS, and far fewer measurements were taken with the iPhone,
I will not list these measurements. They do provide a consistency
check on the Garmin GPS results, and indeed the iPhone coordinates
appeared consistent with the Garmin results at the $\sim$1'' level,
when available.

\begin{figure*}[htp!]
\epsscale{1.0}
\plotone{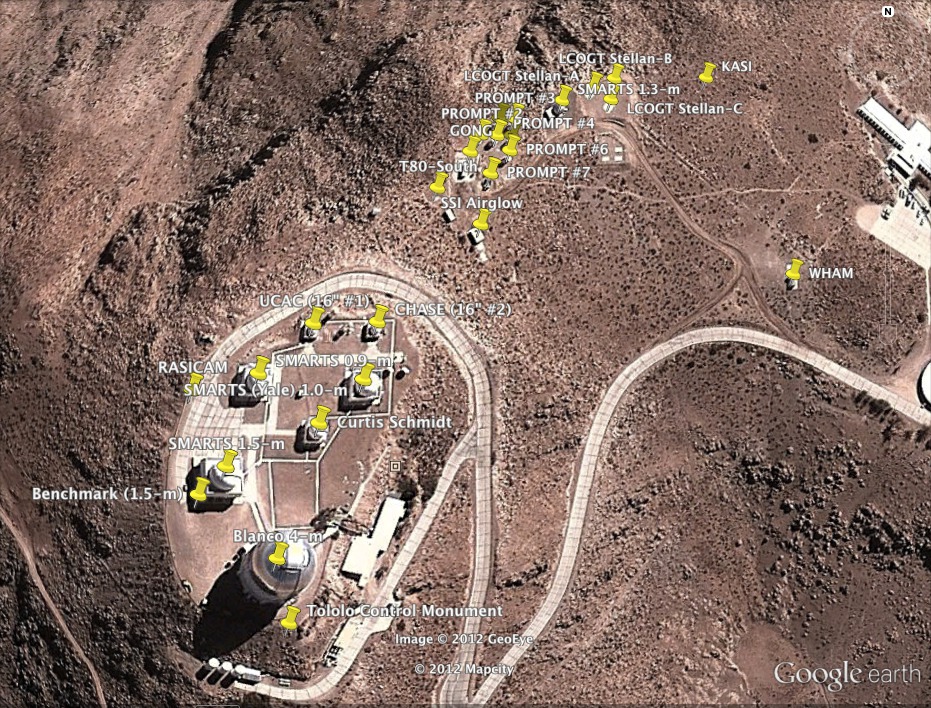}
\caption{Google Earth imagery of Cerro Tololo with observatories and
survey monuments marked. 
\label{Tololo}}
\end{figure*}

\begin{figure*}[htp!]
\epsscale{1.0}
\plotone{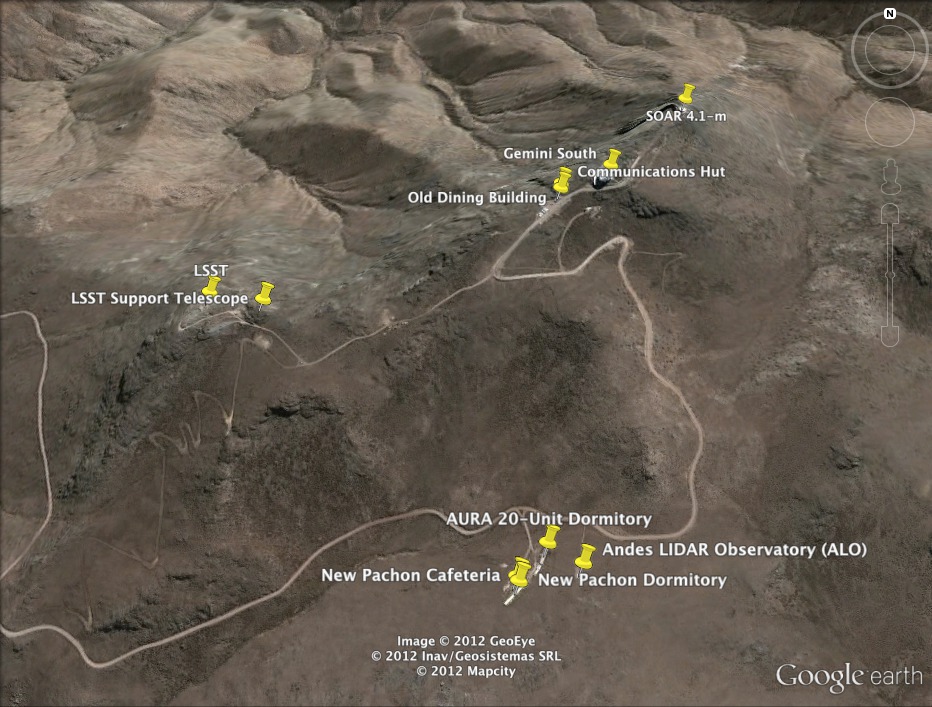}
\caption{Google Earth imagery of Cerro Pach\'{o}n with observatories
and buildings marked. 
\label{Pachon}}
\end{figure*}

{\it Google Earth:} I used Google Earth 5.1 downloaded from the Google
website\footnote{\href{http://www.google.com/earth/index.html}{http://www.google.com/earth/index.html}}
and installed on a MacBook Pro running Mac OS X 10.6.8.  Google Earth
is a ``virtual globe'' that superposes satellite imagery over digital
elevation model (DEM) data from the Shuttle Radar Topography Mission
(SRTM). SRTM was a dual radar system that flew on Space Shuttle
Endeavour in February 2000, producing a high resolution elevation
model, and was a project lead by the National Geospatial-Intelligence
Agency (NGA) and NASA \citep{Farr07}.  SRTM elevation data are on the
WGS~84 coordinate system, adopt the EGM96 vertical datum (geoid), and
for Chile the spatial resolution is 3'' ($\sim$100 m). Hence, the
Google Earth data from SRTM suffers from low resolution which may miss
fine structure, and any leveling of sites over the past decade (e.g.
LSST sites). Google Earth images of Cerro Tololo and Cerro Pach\'{o}n
are shown in Figures \ref{Tololo} and \ref{Pachon}, respectively, with
observatories and survey monuments marked.

\section{Results} 

Individual determinations of the geodetic positions for the
observatories on Cerro Tololo and Cerro Pach\'{o}n are tabulated in Table
\ref{tab:Tololo}. Additional measurements for other facilities on the
two peaks are tabulated in Table \ref{tab:other}.

\begin{deluxetable*}{lllllllll}
\tabletypesize{\scriptsize}
\tablecaption{Measured Geodetic Positions\label{tab:Tololo}}
\tablewidth{0pt}
\tablehead{
\colhead{Site} & \colhead{Method} & \multicolumn{3}{c}{$\phi$} & \multicolumn{3}{c}{$\lambda$}      & \colhead{$H$}\\ 
\colhead{} & \colhead{} & \colhead{$\circ$} & \colhead{'} & \colhead{''} & \colhead{$\circ$} & \colhead{'} & \colhead{''} & \colhead{(m)}
}
\startdata
\hline
Blanco 4-m            & GPS & -30 & 10 & 10.73 & -70 & 48 & 23.52 & 2213\\
(fmr. 150'' or 158'') & GE  & -30 & 10 & 10.73 & -70 & 48 & 23.32 & 2204\\
\hline
SMARTS 1.5-m         & GPS & -30 & 10 & 09.53 & -70 & 48 & 24.49 & 2218\\
(former 60'')        & GE  & -30 & 10 & 09.20 & -70 & 48 & 24.24 & 2208\\
\hline
SMARTS 1.0-m         & GPS & -30 & 10 & 07.94 & -70 & 48 & 21.82 & 2217\\
(former Yale 40'')   & GE  & -30 & 10 & 07.79 & -70 & 48 & 21.69 & 2201\\
\hline
SMARTS 0.9-m         & GPS & -30 & 10 & 08.00 & -70 & 48 & 23.90 & 2217\\
(former 36'')        & GE  & -30 & 10 & 07.69 & -70 & 48 & 23.67 & 2203\\ 
\hline
Curtis Schmidt 0.6-m & GPS & -30 & 10 & 08.59 & -70 & 48 & 22.62 & 2219\\
(former 24'')        & GE  & -30 & 10 & 08.50 & -70 & 48 & 22.49 & 2207\\
\hline
Former UCAC          & GPS & -30 & 10 & 06.93 & -70 & 48 & 22.85 & 2222\\
(former 16'' \#1)    & GE  & -30 & 10 & 06.87 & -70 & 48 & 22.65 & 2200\\
\hline               
Former CHASE            & GPS & -30 & 10 & 06.99 & -70 & 48 & 21.71 & 2217\\
(former 16'' \#2, MMWT) & GE  & -30 & 10 & 06.84 & -70 & 48 & 21.44 & 2196\\
\hline
RASICAM              & GPS & -30 & 10 & 07.99 & -70 & 48 & 25.10 & 2214\\
...                  & GE  & -30 & 10 & 07.96 & -70 & 48 & 24.97 & 2200\\
\hline 
WHAM                 & GPS & -30 & 10 & 05.84 & -70 & 48 & 12.79 & 2159\\
...                  & GE  & -30 & 10 & 05.89 & -70 & 48 & 12.77 & 2154\\
\hline
KASI                 & GPS & -30 & 10 & 01.84 & -70 & 48 & 14.39 & 2157\\
...                  & GE  & ... & ...& ...   & ... & ...& ...   & 2140$^a$\\
\hline
LCOGT Stellan-A      & GPS & -30 & 10 & 02.67 & -70 & 48 & 17.33 & 2170\\
...                  & GE  & -30 & 10 & 02.38 & -70 & 48 & 17.00 & 2146\\
\hline
LCOGT Stellan-B      & GPS & -30 & 10 & 02.48 & -70 & 48 & 16.90 & 2170\\
...                  & GE  & -30 & 10 & 02.20 & -70 & 48 & 16.52 & 2144\\
\hline 
LCOGT Stellan-C      & GPS & -30 & 10 & 02.87 & -70 & 48 & 16.93 & 2170\\
...                  & GE  & -30 & 10 & 02.65 & -70 & 48 & 16.63 & 2151\\
\hline 
SMARTS 1.3-m         & GPS & -30 & 10 & 02.84 & -70 & 48 & 18.01 & 2170\\
...                  & GE  & -30 & 10 & 02.67 & -70 & 48 & 17.68 & 2151\\
\hline 
PROMPT \#1           & GE  & -30 & 10 & 03.42 & -70 & 48 & 18.77 & 2162\\
\hline
PROMPT \#2           & GE  & -30 & 10 & 03.39 & -70 & 48 & 19.42 & 2161\\
\hline
PROMPT \#3           & GE  & -30 & 10 & 03.07 & -70 & 48 & 18.71 & 2157\\
\hline 
PROMPT \#4           & GPS & -30 & 10 & 03.59 & -70 & 48 & 19.40 & 2178\\
...                  & GE  & -30 & 10 & 03.42 & -70 & 48 & 19.09 & 2162\\ 
\hline
PROMPT \#5           & GE  & -30 & 10 & 03.06 & -70 & 48 & 19.02 & 2156\\
\hline
PROMPT \#6           & GE  & -30 & 10 & 03.72 & -70 & 48 & 18.82 & 2166\\
\hline
PROMPT \#7           & GPS & -30 & 10 & 04.38 & -70 & 48 & 19.34 & 2179\\
...                  & GE  & -30 & 10 & 04.17 & -70 & 48 & 19.23 & 2172\\
\hline
GONG                 & GPS & -30 & 10 & 03.98 & -70 & 48 & 19.89 & 2184\\
...                  & GE  & -30 & 10 & 03.75 & -70 & 48 & 19.64 & 2167\\
\hline
SSI Airglow          & GPS & -30 & 10 & 05.35 & -70 & 48 & 19.55 & 2183\\
...                  & GE  & -30 & 10 & 05.11 & -70 & 48 & 19.43 & 2178\\
\hline
SARA South 0.6-m     & GPS & -30 & 10 & 19.73 & -70 & 47 & 57.11 & 2123\\
(Lowell 24'')        & GE  & -30 & 10 & 19.61 & -70 & 47 & 57.00 & 2112\\
\hline 
T80-South (site)     & GPS & -30 & 10 & 04.31 & -70 & 48 & 20.48 & 2187\\
...                  & GE  & ... & ... & ...  & ... & ...& ...   & 2175$^a$\\
\hline 
Gemini South           & GPS & -30 & 14 & 26.54 & -70 & 44 & 12.13 & 2722\\
...                    & GE  & -30 & 14 & 26.70 & -70 & 44 & 11.84 & 2711\\
\hline
SOAR                   & GPS & -30 & 14 & 16.21 & -70 & 44 & 00.84 & 2712\\
...                    & GE  & -30 & 14 & 16.51 & -70 & 44 & 01.24 & 2688\\
\hline
LSST 8.4-m (site)      & GPS & -30 & 14 & 40.68 & -70 & 44 & 57.90 & 2652 \\
...                    & GE  & ... & ... & ...  & ... & ...& ...   & 2633$^a$\\
\hline 
LSST Auxiliary 1.4-m (site) & GPS & -30 & 14 & 41.27 & -70 & 44 & 51.80 & 2652\\
...                    & GE  & ... & ... & ...  & ... & ...& ...   & 2627$^a$\\
\hline
Andes LIDAR Obs. (ALO) & GPS & -30 & 15 & 06.39 & -70 & 44 & 17.48 & 2519\\
...                    & GE  & -30 & 15 & 06.24 & -70 & 44 & 17.37 & 2522
\enddata
\tablecomments{GPS = Global Positioning System, measured using Garmin
  Dakota 20 unit. GE = Google Earth (26 March 2011 image of Tololo, 11
  Apr 2011 image of Pach\'{o}n). Geodetic latitude $\phi$, longitude
  $\lambda$ on WGS~84 system. Elevation $H$ is orthometric height with
  respect to geoid. Google~Earth adopts the EGM96 geoid, however the
  adopted geoid used by the Dakota~20 GPS is ambiguous.  The KASI,
  T80-S, LSST, and LSST Support sites were undeveloped at the time of
  the latest Google Earth images (hence no coordinates could be
  determined visually). $a$ = Google Earth elevation at the GPS
  position for this facility. The LSST and LSST Auxiliary sites were
  unleveled during the time ($\sim$2000) of the SRTM elevation mapping
  used by Google~Earth, so are not reliable.}
\end{deluxetable*}

\begin{deluxetable*}{llllllllll}
\tabletypesize{\scriptsize}
\tablecaption{Measured Geodetic Positions for Other Landmarks\label{tab:other}}
\tablewidth{0pt}
\tablehead{
\colhead{Site} & \colhead{Method} & \multicolumn{3}{c}{$\phi$} & \multicolumn{3}{c}{$\lambda$}      & \colhead{$H$}\\ 
\colhead{} & \colhead{} & \colhead{$\circ$} & \colhead{'} & \colhead{''} & \colhead{$\circ$} & \colhead{'} & \colhead{''} & \colhead{(m)}
}
\startdata
\hline 
\href{https://maps.google.com/maps?&z=18&q=-30.172477500,-70.800026667}{NASA Monument 7401}     & GPS  & -30 & 10 & 20.919   & -70 & 48 & 00.096   & 2124.1\\
(Site 892, W of SARA)  & GE   & -30 & 10 & 20.90    & -70 & 48 & 00.01    & 2124\\ 
\hline 
\href{https://maps.google.com/maps?&z=18&q=-30.169377778,-70.806925000}{Monument (SW of 1.5-m)} & GPS & -30 & 10 & 09.76     & -70 & 48 & 24.93    & 2214.0\\
...                    & GE  & -30 & 10 & 09.67     & -70 & 48 & 24.78    & 2206\\
\hline
\href{https://maps.google.com/maps?&z=18&q=-30.1699305556,-70.806458333}{Tololo Control Monument} & GPS & -30 & 10 & 11.75    & -70 & 48 & 23.25    & 2222\\
(SE of Blanco)          & GE  & ... & ... & ...     & ... & ... & ... & 2201\\
\hline
\href{https://maps.google.com/maps?&z=18&q=-30.168863888,-70.802647222}{Round Office Building} & GPS & -30 & 10 & 07.91 & -70 & 48 & 09.53 & 2158\\
(Tololo)               & GE  & -30 & 10 & 07.83 & -70 & 48 & 09.40 & 2151\\
\hline
\href{https://maps.google.com/maps?&z=18&q=-30.251216667,-70.7391111}{AURA 20 Unit Dormitory} & GPS & -30 & 15 & 04.38 & -70 & 44 & 20.80 & 2502\\
(Pach\'{o}n)               & GE  & -30 & 15 & 04.34 & -70 & 44 & 20.75 & 2506\\
\hline
\href{https://maps.google.com/maps?&z=18&q=-30.252094444,-70.73990000}{New Dormitory}       & GPS & -30 & 15 & 07.54 & -70 & 44 & 23.64 & 2503\\
(Pach\'{o}n)               & GE  & -30 & 15 & 07.64 & -70 & 44 & 23.80 & 2503\\
\hline
\href{https://maps.google.com/maps?&z=18&q=-30.252016667,-70.7400027778}{New Dining Building} & GPS & -30 & 15 & 07.26 & -70 & 44 & 24.01 & 2503\\
(Pach\'{o}n)               & GE  & -30 & 15 & 07.22 & -70 & 44 & 23.97 & 2500\\ 
\hline
\href{https://maps.google.com/maps?&z=18&q=-30.241458333,-70.73826667}{Old Dining Building} & GPS & -30 & 14 & 29.25 & -70 & 44 & 17.76 & 2699\\
(Pach\'{o}n)               & GE  & -30 & 14 & 29.28 & -70 & 44 & 17.90 & 2692\\
\hline
\href{https://maps.google.com/maps?&z=18&q=-30.2412916667,-70.73826667}{Communications Hut} & GPS & -30 & 14 & 28.65 & -70 & 44 & 17.76 & 2702\\ 
(Pach\'{o}n)               & GE  & -30 & 14 & 28.69 & -70 & 44 & 17.80 & 2692\\
\hline
\href{https://maps.google.com/maps?&z=18&q=-29.917019444,-71.241947222}{CTIO Office Entrance} & GPS & -29 & 55 & 01.27 & -71 & 14 & 31.01 &  97\\
(La Serena)            & GE  & -29 & 55 & 01.37 & -71 & 14 & 30.97 &  90\\
\hline                                       
\href{https://maps.google.com/maps?&z=18&q=-29.916527778,-71.2417527778}{Gemini Office Entrance} & GPS & -29 & 54 & 59.50 & -71 & 14 & 30.31 &  97\\
(La Serena)            & GE  & -29 & 54 & 59.65 & -71 & 14 & 30.44 &  94\\
\hline                                       
\href{https://maps.google.com/maps?&z=18&q=-29.91732058,-71.24217972}{SOAR Office Entrance} & GPS & -29 & 55 & 02.35 & -71 & 14 & 31.85 &  97\\
(La Serena)            & GE  & -29 & 55 & 02.38 & -71 & 14 & 31.21 &  88
\enddata
\tablecomments{GPS = Global Positioning System, measured using Garmin
  Dakota 20 unit. GE = Google Earth. Geodetic latitude $\phi$,
  longitude $\lambda$ on WGS~84 system. Elevation $H$ is orthometric
  height with respect to geoid.}
\end{deluxetable*}

\subsection{Comparison of Google~Earth to GPS \label{comparison}}

For approximately 30 structures (observatories, buildings, and
monuments) on Cerro Tololo and Cerro Pach\'{o}n for which I measured both
positions via GPS and Google~Earth (GE), I find the following offsets
in geodetic latitude $\phi$ and longitude $\lambda$:

\begin{equation}
\phi_{GE}\,-\,\phi_{GPS}\,=\,-0".10\,\pm\,0".03\,({\rm rms} = 0".14) 
\end{equation}

\begin{equation}
\lambda_{GE}\,-\,\lambda_{GPS}\,=\,-0".14\,\pm\,0".02\,({\rm rms} = 0".14) 
\end{equation}

in units of length, this translates to:

\begin{equation}
\phi_{GE}\,-\,\phi_{GPS}\,=\,-3.1\,\pm\,1.0\,({\rm rms} = 4.3)~{\rm m}
\end{equation}

\begin{equation}
\lambda_{GE}\,-\,\lambda_{GPS}\,=\,-3.8\,\pm\,0.6\,({\rm rms} = 3.8)~{\rm m}
\end{equation}

When comparing Google Earth imagery at different epochs, systematic
shifts in latitude and longitude are visible.  Repeated measurements
of the position for the NASA geodetic monument (just west of the SARA
South observatory) on Google~Earth at different dates (27 Feb 2006, 3
Apr 2010, 26 Mar 2011) show systematic epoch-to-epoch shifts at the
$\pm$0''.08-0''.13 level. Hence, for the purposes of deriving ``best''
geodetic positions, I correct the Google~Earth positions to take into
account their systematic difference with respect to GPS positions.

\subsection{NASA Geodetic Monument \label{NASA_Monument}}

\begin{figure}
\epsscale{1.0}
\plotone{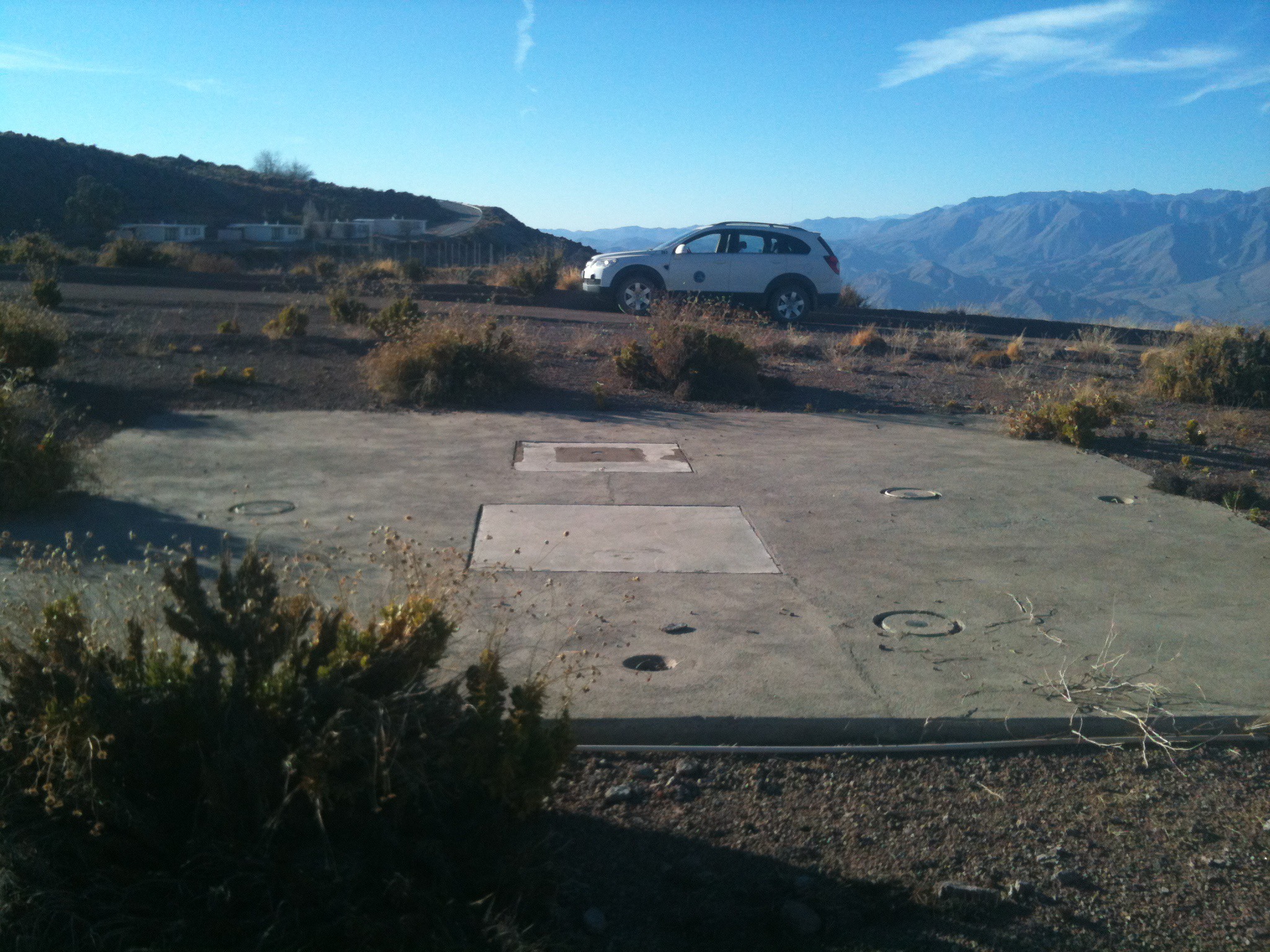}
\caption{Platform for monument 7401 (site \#892) situated just west of the
SARA South Observatory (former Lowell 24''). The benchmark was 
built for the NASA Crustal Dynamics Project in the early 1980's, and 
satellite laser ranging measurements were taken here between
1984 and 1991. The actual benchmark is marked by a metal disk within
the central dark square on the platform. Several auxiliary benchmarks 
labeled with 7401 and letters are visible scattered across the platform.
The SARA South Observatory lies to the right of the image down the
service road that the car is parked on.
\label{NASA7401}}
\end{figure}

The NASA Crustal Dynamics Project established a cluster of observing
monuments near the SARA South Observatory (former Lowell 24'') for
satellite laser ranging measurements \citep[the technique is discussed
  in e.g.][]{Tapley85}. The monuments are collectively referred to as
``site number 892''\footnote{Data is provided at the website
  \href{http://cddis.nasa.gov/site$\_$cat/cerr.html}{http://cddis.nasa.gov/site$\_$cat/cerr.html}
  maintained by the Space Geodesy and Altimetry Projects Office
  (SGAPO), edited by Mark Bryant and Carey Noll, dated March 1993.},
but the primary monument is labeled number 7401, and it occasionally
appears by this number in the geodesy literature. The monument disk is
about a meter north of the center of a 25 foot square pad easily
visible on Google Earth (and easily visible just south of the SARA
telescope access road), and shown in Figure \ref{NASA7401}. The site
hosted three campaigns using Transportable Laser Ranging Systems
(TLRS) between 1984 and 1991 \citep[the program is discussed
  in][]{Allenby84}, taking range measurements to the Laser Geodynamics
Satellite (LAGEOS).  Precise geodetic coordinates for monument 7401
are provided in the SGAPO online archives as
\href{https://maps.google.com/maps?&z=18&q=-30.172463033333,-70.800042677778}{latitude
  south 30$^{\circ}$ 10' 20.86692'', longitude west 70$^{\circ}$ 48'
  00.15364''}, elevation 2123.090 m, height above ellipsoid 2155.748 m
(dated 23 April 1990).  The quoted ellipsoid assumed an equatorial
radius of 6378137 m and flattening f = 1/298.255, i.e. an identical
radius, but slightly different flattening compared to WGS~84.

The monument provides a useful check on the types of ``elevations''
that are being reported. Both GPS and Google~Earth measured an
elevation of 2124 m for the NASA monument, i.e. only 1 m above the
NASA geodetic value. This measurement alone is strongly suggestive
that both the GPS and Google~Earth measurements take into account a
geoid model, and are not simply measured from the WGS~84 reference
ellipsoid. Unfortunately, the agreement between the GPS and
Google~Earth elevations for the monument does not provide a useful
explanation for the $\sim$11 m systematic offset between the two as
inferred from averages of elevation measures for $\sim$30 other sites.
The source of the discrepancy at other sites is unclear. Is standing
next to buildings biasing the GPS elevations? Is small scale elevation
structure not taken into account by the smoothed SRTM elevation data
affecting the Google Earth elevations?

Preliminary estimates of the motion of monument 7401 were reported in
\citet{Smith94}, however the errors were large. The measured motion
during 1984-1991 was 35.9 mm/yr towards azimuth angle 30$^{\circ}$,
however the error ellipsoid was 24.0 mm/yr $\times$ 10.7 mm/yr with
the long axis oriented towards azimuth angle -5$^{\circ}$. Hence, the
measured motion was statistically negligible. The NASA geodetic
coordinates are valuable, however, as they provide not only a
well-calibrated position for comparison with the GPS and Google Earth
positions, but a potential first epoch for estimating the motion of
Cerro Tololo with respect to the WGS~84 terrestrial coordinate system.

\subsubsection{Tololo Control Monument \label{Tololo_Monument}}

\begin{figure}
\epsscale{1.0}
\plotone{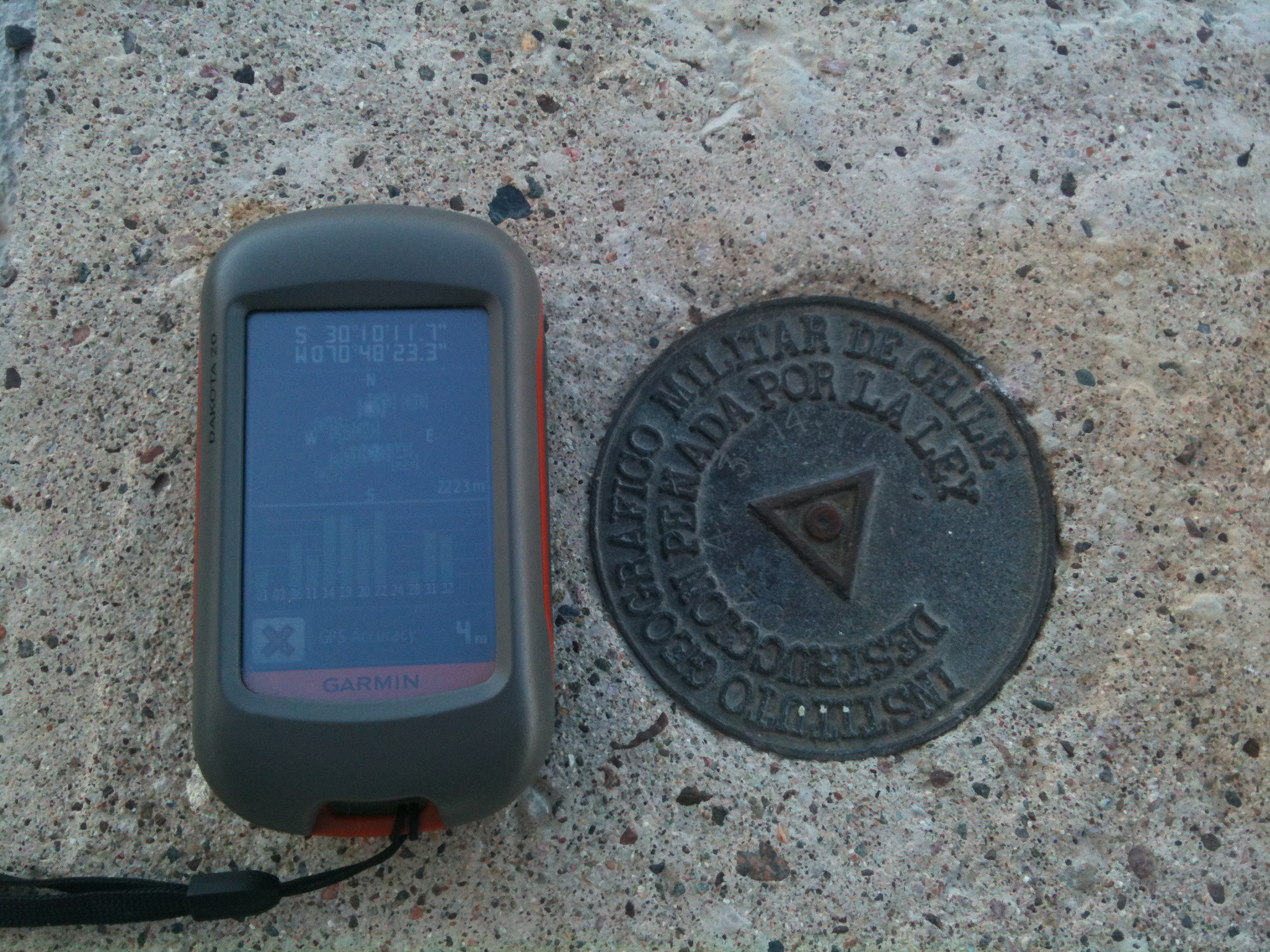}
\caption{``Tololo Control'' survey monument on the Cerro Tololo peak,
  just southeast of the Blanco 4-m telescope. This was the primary
  benchmark for Tololo in the surveys of the mid-1960s, and its
  elevation is within $\sim$1-2 m of the platforms for all of the
  telescopes on the Tololo plateau. The 2008 survey by Juan Carlos
  Aravena Godoy estimated a geodetic position of $\phi$ =
  -30$^{\circ}$10'11.7492'', $\lambda$ = -70$^{\circ}$48'23.2709'' at
  elevation 2242.493\,m above the GRS~80 ellipsoid (i.e. not taking
  into account a geoid model, not with respect to mean sea level).
\label{TololoMonument}}
\end{figure}

A survey monument called
\href{https://maps.google.com/maps?&z=18&q=-30.1699303333,-70.8064641388889}{``Tololo'',
  ``Tololo Control'', or ``O''} on the 1960s/1970s-era notes and
survey maps of Cerro Tololo lies just southeast of the Blanco 4-m dome
structure, just outside of the guard rail girding the gravel (see Fig.
\ref{TololoMonument}).  In Table \ref{tab:TololoControl}, I list
several estimates of the elevation of the
\href{https://maps.google.com/maps?&z=18&q=-30.1699303333,-70.8064641388889}{Tololo
  Control survey monument}. An elevation of 2211.60 m above sea level
is listed on a surveying note from the Kitt Peak National Observatory
Engineering Dept. This benchmark was used in conjunction with
monuments on other nearby peaks (Pach\'{o}n, Morado, etc.) for tying
in the position of Tololo with the cartographic grid used by the
Instituto Geographico Militar de Chile and Chilean Air Force in the
1960s (the Chile Plane Coordinate System)\footnote{The monument disk
  states ``Instituto Geographico Militar de Chile - destruccion penada
  por la ley''.}. The elevation of this monument defined the elevation
scale for other structures built on the Tololo plateau, as marked in
the 1973 survey map (see Table \ref{tab:elev}).  GPS measurement of
the elevation of the
\href{https://maps.google.com/maps?&z=18&q=-30.1699303333,-70.8064641388889}{Tololo
  monument} yielded 2222 m. The 2008 survey discussed in
\S\ref{2008survey} measured a geodetic elevation of 2242.493 m with
respect to the GRS~80 ellipsoid (which is within tens of micrometers
of the WGS~84 ellipsoid).  The monument is not obvious on Google~Earth
imagery, however the elevation in the vicinity of the monument is 2201
m.

\begin{deluxetable*}{lllll}
\tabletypesize{\scriptsize}
\tablecaption{Elevations for Tololo Control Monument\label{tab:TololoControl}}
\tablewidth{0pt}
\tablehead{
\colhead{Source}      & \colhead{Elevation} & \colhead{Notes}\\
\colhead{}            & \colhead{(m)}       & \colhead{}}
\startdata
1973 survey           & 2211.60\,m        & ``mean sea level'' \\
{\bf 2008 survey}     & {\bf 2242.493\,m} & w/r GRS~80 ellipsoid (no geoid)\\
Google Earth          & 2201\,m           & w/r WGS~84 ellipsoid + EGM96 geoid\\
GPS                   & 2222\,m           & w/r WGS~84 ellipsoid + unknown geoid
\enddata
\end{deluxetable*}
\vspace{0.5cm}

The 2008 survey measurement for the Tololo Control survey marker is
the most accurate WGS~84 coordinate position, and fortunately its
datum is unambiguous (GRS~80 $\approx$ WGS~84). The GPS elevation is
20\,m lower than the 2008 geodetic survey elevation.  The Google~Earth
elevation is $\sim$10 m lower yet, however Google~Earth measures
elevations with respect to the EGM-96 geoid, which accounts for most
of the $\sim$41\,m discrepancy.  The level of disagreement among the
elevations is somewhat surprising, however, given the good agreement
for the elevations measured for the NASA monument near the SARA South
observatory.

\subsubsection{Treatment of Elevation Data}

Multiple series of GPS measurements were taken of a few sites on
different days. From comparing average GPS elevation measurements
taken on different days, it is clear that significant systematic
errors in the GPS elevations are present at the $\pm$5 m rms level
(presumably from having data from different satellites participate in
the solution).  This precludes using the standard error of the mean
for measurements taken during a short interval on a given day as a
useful measure of the uncertainties.

Inter-comparison of the GPS and Google~Earth elevations reveals a
systematic offset: 

\begin{equation}
  <H_{Google} - H_{GPS}> = -11.5\,\pm\,2.0~{\rm m}
\end{equation}

The rms scatter was $\pm$8.6 meters. The source of the discrepancy is
unclear. Two likely possibilities are that either (1) the 3''
resolution of the Google Earth elevation data from SRTM has smoothed
over the high elevation points (where many of the telescopes are
located), or (2) there may be a difference in the geoid adopted in the
GPS calculations and that used by Google Earth (EGM96). Both the GPS
and Google~Earth elevations are clearly limited in their accuracy.
The GPS measurements often have epoch-to-epoch jumps and the datum is
ambiguous (at the time of writing, there is conflicting literature on
the web regarding this; whether Garmin GPSs measure with respect to
the ellipsoid, or whether they include a geoid model). The Google
Earth maps have excellent imagery resolution, but the elevation maps
are limited to the SRTM resolution ($\sim$3''). Unfortunately,
elevations taken both with the handheld GPS unit and via Google Earth
must both be taken with a grain of salt. The best elevations that we
have available are those from the 2008 survey and those for a NASA
geodetic monument.

For useful comparison of the elevations, it was decided to place all
measurements on the same elevation system defined by the 2008
theodolite survey (which includes very accurate elevations measured
with respect to the GRS~80 [$\approx$WGS~84] ellipsoid). Through
inter-comparison of the measured elevations, I decided to correct GPS,
Google Earth, and 1973 survey elevations to the geodetic elevation
system defined by the 2008 survey (close enough to WGS~84 to be
practically considered WGS~84):

\begin{equation}
h_{WGS~84}~\simeq~H_{GPS} + (28\,\pm\,3)~{\rm m} 
\end{equation}

\begin{equation}
h_{WGS~84}~\simeq~H_{Google~Earth} + N(EGM96)
\end{equation}

\begin{equation}
h_{WGS~84}~\simeq~H_{1973~Survey} + (31.3\,\pm\,0.4)~{\rm m}
\end{equation}

The difference between the Google~Earth elevations and the 2008 survey
geodetic heights $h$ are close enough to the EGM-96 geoid undulations
$N$ (38\,$\pm$\,3 m; see actual EGM-96 $N$ values in Table 6), that
they appear to provide useful elevations $H$. The difference between
the GPS elevations and 2008 survey heights do not appear to correspond
to either zero (if the GPS elevations were actually geodetic heights),
or any of the widely used geoid undulations. Hence, I simply add a
constant to the GPS elevations and place them on the WGS~84 geodetic
height scale. The difference between the 2008 survey geodetic heights
and the 1973 survey elevations (31.3 m) is remarkably similar to the
geoid undulation heights (\S2; especially for the WGS~84 ellipsoid),
suggesting that the zero point for the 1973 survey elevations is
indeed within a few meters of recent geoid models.

\section{Conclusions}

Best estimates for the geodetic and geocentric positions of the
observatories on Cerro Tololo and Cerro Pach\'{o}n are listed in Table
\ref{tab:final}. Final elevations taking into account the offsets in
elevations discussed in the last section are also included, along
with geoid undulation heights $N$ based on the EGM-96 model.

It appears that using Google Earth one can derive geodetic WGS~84
coordinates for observatories and structures to approximately
$\pm$0''.15 (5~m) accuracy in latitude and longitude without resorting
to more accurate techniques (i.e. GPS, surveying).  Google~Earth shows
systematic shifts in position between imagery epochs at the
$\sim$$\pm$0''.1 level ($\sim$3~m). Positions measured on the most
recent Google~Earth imagery of Cerro Tololo and Pach\'{o}n reveal slight
systematic offsets with respect to GPS-derived positions at the
$\pm$0''.1 (3~m) level, which can be corrected for if necessary. The
rms scatter in positions derived via GPS and Google~Earth is
approximately 0''.14 (4~m).  Treating the GPS and Google~Earth
positions as independent estimates, and correcting the Google~Earth
positions for small systematic differences at the 0''.1 (3~m) level,
one can derive final mean positions for structures on Cerro Tololo and
Pach\'{o}n with absolute accuracy $\pm$0''.1 (3~m).  One can obviously do
better if needed through long integrations of GPS determinations, or
theodolite observations tied to monuments with well-constrained
geodetic positions, however this would obviously require more time and
money (which always seem to be in short supply).

While {\it astronomical} coordinates have not been determined for all
of the observatories on Cerro Tololo and Cerro Pach\'{o}n (an arduous
task), one can derive approximate astronomical coordinates by adding
an offset derived from Harrington et al.'s (1972) observations:

\begin{equation}
\phi_{astronomical}~-~\phi_{geodetic}~\simeq\,12".7\,\pm\,1".5
\end{equation}

\begin{equation}
\lambda_{astronomical}~-~\lambda_{geodetic}~\simeq\,-31".6\,\pm\,2".0
\end{equation} 

However this is only a rough approximation as no doubt the vertical
deflection due to the gravity field varies subtly over the two peaks.

Besides the obvious (and explainable) disagreement between the
geodetic and astronomical positions for Cerro Tololo (\S\ref{1972}),
there was at least one other surprise in comparing the final positions
in Table \ref{tab:final} with the previously published positions in
Table \ref{tab:old}. An often quoted
\href{https://maps.google.com/maps?&z=18&q=-30.17225000,-70.800027778}{position
  for LSST} listed in two recent LSST documents
\citep{Ivezic08B,Ivezic08} appears to be in error by 9.4 km
($\Delta$$\phi$ = 260".6, $\Delta$$\lambda$ = 60.0''), and by
$\sim$500 m in elevation (2123 m listed vs.  $\sim$2600 m
measured). The position listed in \citet{Ivezic08B} and
\citet{Ivezic08} appears to correspond most closely to the
\href{https://maps.google.com/maps?&z=18&q=-30.1724630333,-70.80004267778}{NASA
  survey monument on Cerro {\it Tololo}} (rather than Pach\'{o}n).
The position for the LCOGT observatories on Tololo on the LCOGT
website appear to be in error by $\sim$0.2 km. Our position for Gemini
South agrees with the value on their
website\footnote{\href{http://www.gemini.edu/sciops/telescopes-and-sites/locations}{http://www.gemini.edu/sciops/telescopes-and-sites/locations}}
to better than $<$0''.04 for each axis (within $\sim$1 m), and our GPS
estimate of the elevation (2722 m) is identical to their value.

After finding that some of the published positions for observatories
on Cerro Tololo and Cerro Pach\'{o}n were erroneous, it may be worth
the time to double-check the published positions of other
observatories (especially if those observatories pre-date the GPS
era). Using erroneous positions for observatories can lead to
unnecessary systematic errors in various position-dependent
astronomical calculations, which are inexcusable in the era of GPS and
Google Earth.

\acknowledgements

The author thanks Jason Wright for pointing out the discrepancy
between the published position for CTIO and that determined with
Google Earth, which triggered this investigation. I thank Michael
Warner for lending the author his Garmin GPS. The author also thanks
Enrique Figueroa for sharing the 2008 survey report from Juan Carlos
Aravena Godoy, and thanks to Malcolm Smith, Oscar Saa, Tim Abbott,
Alistair Walker, Chris Smith, Nicole van der Bliek, Kadur Flores, Jeff
Barr, and Dan Phillips, for conversations on the history and mythology
of Cerro Tololo and Cerro Pach\'{o}n.


\begin{thebibliography}{26}
\expandafter\ifx\csname natexlab\endcsname\relax\def\natexlab#1{#1}\fi

\bibitem[{{Allenby}(1984)}]{Allenby84}
{Allenby}, R.~J. 1984, {Andean tectonics: Implications for Satellite Geodesy},
  Tech. rep.

\bibitem[{{Blanco} \& {Mayall}(1972)}]{Blanco72}
{Blanco}, V.~M. \& {Mayall}, N.~U. 1972, in Bulletin of the American
  Astronomical Society, Vol.~4, Bulletin of the American Astronomical Society,
  66--97

\bibitem[{{Buie} {et~al.}(2012){Buie}, {Wasserman}, {Vaduvescu}, {Holmes}, \&
  {Linder}}]{Buie12}
{Buie}, M.~W., {Wasserman}, L.~H., {Vaduvescu}, O., {Holmes}, R., \& {Linder},
  T. 2012, Minor Planet Circulars, 79528, 6

\bibitem[{{Elliot} {et~al.}(1981){Elliot}, {Frogel}, {Elias}, {Glass},
  {French}, {Mink}, \& {Liller}}]{Elliot81}
{Elliot}, J.~L., {Frogel}, J.~A., {Elias}, J.~H., {Glass}, I.~S., {French},
  R.~G., {Mink}, D.~J., \& {Liller}, W. 1981, \aj, 86, 127

\bibitem[{{Farr} {et~al.}(2007){Farr}, {Rosen}, {Caro}, {Crippen}, {Duren},
  {Hensley}, {Kobrick}, {Paller}, {Rodriguez}, {Roth}, {Seal}, {Shaffer},
  {Shimada}, {Umland}, {Werner}, {Oskin}, {Burbank}, \& {Alsdorf}}]{Farr07}
{Farr}, T.~G., {Rosen}, P.~A., {Caro}, E., {Crippen}, R., {Duren}, R.,
  {Hensley}, S., {Kobrick}, M., {Paller}, M., {Rodriguez}, E., {Roth}, L.,
  {Seal}, D., {Shaffer}, S., {Shimada}, J., {Umland}, J., {Werner}, M.,
  {Oskin}, M., {Burbank}, D., \& {Alsdorf}, D. 2007, Reviews of Geophysics, 45,
  2004

\bibitem[{{French} {et~al.}(1983){French}, {Elias}, {Mink}, \&
  {Elliot}}]{French83}
{French}, R.~G., {Elias}, J.~H., {Mink}, D.~J., \& {Elliot}, J.~L. 1983,
  \icarus, 55, 332

\bibitem[{{Hamuy} {et~al.}(1996){Hamuy}, {Phillips}, {Suntzeff}, {Schommer},
  {Maza}, \& {Aviles}}]{Hamuy96}
{Hamuy}, M., {Phillips}, M.~M., {Suntzeff}, N.~B., {Schommer}, R.~A., {Maza},
  J., \& {Aviles}, R. 1996, \aj, 112, 2391

\bibitem[{{Harrington} {et~al.}(1972){Harrington}, {Mintz Blanco}, \&
  {Blanco}}]{Harrington72}
{Harrington}, R.~S., {Mintz Blanco}, B., \& {Blanco}, V.~M. 1972,
  {Contributions of the Cerro Tololo Inter-American Observatory No. 126:
  Geodetic and Astronomical Coordinates of the Cerro Tololo Inter-American
  Observatory}

\bibitem[{{Hofmann-Wellenhof} {et~al.}(1992){Hofmann-Wellenhof},
  {Lichtenegger}, \& {Collins}}]{GPS92}
{Hofmann-Wellenhof}, B., {Lichtenegger}, H., \& {Collins}, J. 1992, {Global
  Positioning System (GPS). Theory and practice}

\bibitem[{{Hubbard} {et~al.}(1997){Hubbard}, {Porco}, {Hunten}, {Rieke},
  {Rieke}, {McCarthy}, {Haemmerle}, {Haller}, {McLeod}, {Lebofsky},
  {Marcialis}, {Holberg}, {Landau}, {Carrasco}, {Elias}, {Buie}, {Dunham},
  {Persson}, {Boroson}, {West}, {French}, {Harrington}, {Elliot}, {Forrest},
  {Pipher}, {Stover}, {Brahic}, \& {Grenier}}]{Hubbard97}
{Hubbard}, W.~B., {Porco}, C.~C., {Hunten}, D.~M., {Rieke}, G.~H., {Rieke},
  M.~J., {McCarthy}, D.~W., {Haemmerle}, V., {Haller}, J., {McLeod}, B.,
  {Lebofsky}, L.~A., {Marcialis}, R., {Holberg}, J.~B., {Landau}, R.,
  {Carrasco}, L., {Elias}, J., {Buie}, M.~W., {Dunham}, E.~W., {Persson},
  S.~E., {Boroson}, T., {West}, S., {French}, R.~G., {Harrington}, J.,
  {Elliot}, J.~L., {Forrest}, W.~J., {Pipher}, J.~L., {Stover}, R.~J.,
  {Brahic}, A., \& {Grenier}, I. 1997, \icarus, 130, 404

\bibitem[{{Ivezic} {et~al.}(2008{\natexlab{a}}){Ivezic}, {Axelrod}, {Brandt},
  {Burke}, {Claver}, {Connolly}, {Cook}, {Gee}, {Gilmore}, {Jacoby}, {Jones},
  {Kahn}, {Kantor}, {Krabbendam}, {Lupton}, {Monet}, {Pinto}, {Saha}, {Schalk},
  {Schneider}, {Strauss}, {Stubbs}, {Sweeney}, {Szalay}, {Thaler}, {Tyson}, \&
  {LSST Collaboration}}]{Ivezic08B}
{Ivezic}, Z., {Axelrod}, T., {Brandt}, W.~N., {Burke}, D.~L., {Claver}, C.~F.,
  {Connolly}, A., {Cook}, K.~H., {Gee}, P., {Gilmore}, D.~K., {Jacoby}, S.~H.,
  {Jones}, R.~L., {Kahn}, S.~M., {Kantor}, J.~P., {Krabbendam}, V.~V.,
  {Lupton}, R.~H., {Monet}, D.~G., {Pinto}, P.~A., {Saha}, A., {Schalk}, T.~L.,
  {Schneider}, D.~P., {Strauss}, M.~A., {Stubbs}, C.~W., {Sweeney}, D.,
  {Szalay}, A., {Thaler}, J.~J., {Tyson}, J.~A., \& {LSST Collaboration}.
  2008{\natexlab{a}}, Serbian Astronomical Journal, 176, 1

\bibitem[{{Ivezic} {et~al.}(2008{\natexlab{b}}){Ivezic}, {Tyson}, {Acosta},
  {Allsman}, {Anderson}, {Andrew}, {Angel}, {Axelrod}, {Barr}, {Becker},
  {Becla}, {Beldica}, {Blandford}, {Bloom}, {Borne}, {Brandt}, {Brown},
  {Bullock}, {Burke}, {Chandrasekharan}, {Chesley}, {Claver}, {Connolly},
  {Cook}, {Cooray}, {Covey}, {Cribbs}, {Cutri}, {Daues}, {Delgado}, {Ferguson},
  {Gawiser}, {Geary}, {Gee}, {Geha}, {Gibson}, {Gilmore}, {Gressler}, {Hogan},
  {Huffer}, {Jacoby}, {Jain}, {Jernigan}, {Jones}, {Juric}, {Kahn}, {Kalirai},
  {Kantor}, {Kessler}, {Kirkby}, {Knox}, {Krabbendam}, {Krughoff}, {Kulkarni},
  {Lambert}, {Levine}, {Liang}, {Lim}, {Lupton}, {Marshall}, {Marshall}, {May},
  {Miller}, {Mills}, {Monet}, {Neill}, {Nordby}, {O'Connor}, {Oliver},
  {Olivier}, {Olsen}, {Owen}, {Peterson}, {Petry}, {Pierfederici},
  {Pietrowicz}, {Pike}, {Pinto}, {Plante}, {Radeka}, {Rasmussen}, {Ridgway},
  {Rosing}, {Saha}, {Schalk}, {Schindler}, {Schneider}, {Schumacher}, {Sebag},
  {Seppala}, {Shipsey}, {Silvestri}, {Smith}, {Smith}, {Strauss}, {Stubbs},
  {Sweeney}, {Szalay}, {Thaler}, {Vanden Berk}, {Walkowicz}, {Warner},
  {Willman}, {Wittman}, {Wolff}, {Wood-Vasey}, {Yoachim}, {Zhan}, \& {for the
  LSST Collaboration}}]{Ivezic08}
{Ivezic}, Z., {Tyson}, J.~A., {Acosta}, E., {Allsman}, R., {Anderson}, S.~F.,
  {Andrew}, J., {Angel}, R., {Axelrod}, T., {Barr}, J.~D., {Becker}, A.~C.,
  {Becla}, J., {Beldica}, C., {Blandford}, R.~D., {Bloom}, J.~S., {Borne}, K.,
  {Brandt}, W.~N., {Brown}, M.~E., {Bullock}, J.~S., {Burke}, D.~L.,
  {Chandrasekharan}, S., {Chesley}, S., {Claver}, C.~F., {Connolly}, A.,
  {Cook}, K.~H., {Cooray}, A., {Covey}, K.~R., {Cribbs}, C., {Cutri}, R.,
  {Daues}, G., {Delgado}, F., {Ferguson}, H., {Gawiser}, E., {Geary}, J.~C.,
  {Gee}, P., {Geha}, M., {Gibson}, R.~R., {Gilmore}, D.~K., {Gressler}, W.~J.,
  {Hogan}, C., {Huffer}, M.~E., {Jacoby}, S.~H., {Jain}, B., {Jernigan}, J.~G.,
  {Jones}, R.~L., {Juric}, M., {Kahn}, S.~M., {Kalirai}, J.~S., {Kantor},
  J.~P., {Kessler}, R., {Kirkby}, D., {Knox}, L., {Krabbendam}, V.~L.,
  {Krughoff}, S., {Kulkarni}, S., {Lambert}, R., {Levine}, D., {Liang}, M.,
  {Lim}, K., {Lupton}, R.~H., {Marshall}, P., {Marshall}, S., {May}, M.,
  {Miller}, M., {Mills}, D.~J., {Monet}, D.~G., {Neill}, D.~R., {Nordby}, M.,
  {O'Connor}, P., {Oliver}, J., {Olivier}, S.~S., {Olsen}, K., {Owen}, R.~E.,
  {Peterson}, J.~R., {Petry}, C.~E., {Pierfederici}, F., {Pietrowicz}, S.,
  {Pike}, R., {Pinto}, P.~A., {Plante}, R., {Radeka}, V., {Rasmussen}, A.,
  {Ridgway}, S.~T., {Rosing}, W., {Saha}, A., {Schalk}, T.~L., {Schindler},
  R.~H., {Schneider}, D.~P., {Schumacher}, G., {Sebag}, J., {Seppala}, L.~G.,
  {Shipsey}, I., {Silvestri}, N., {Smith}, J.~A., {Smith}, R.~C., {Strauss},
  M.~A., {Stubbs}, C.~W., {Sweeney}, D., {Szalay}, A., {Thaler}, J.~J., {Vanden
  Berk}, D., {Walkowicz}, L., {Warner}, M., {Willman}, B., {Wittman}, D.,
  {Wolff}, S.~C., {Wood-Vasey}, W.~M., {Yoachim}, P., {Zhan}, H., \& {for the
  LSST Collaboration}. 2008{\natexlab{b}}, ArXiv e-prints

\bibitem[{{Lasker} {et~al.}(1973){Lasker}, {Bracker}, \& {Kunkel}}]{Lasker73}
{Lasker}, B.~M., {Bracker}, S.~B., \& {Kunkel}, W.~E. 1973, \pasp, 85, 109

\bibitem[{{Mayall}(1968)}]{Mayall68}
{Mayall}, N.~U. 1968, \pasp, 80, 37

\bibitem[{{Perlmutter} {et~al.}(1999){Perlmutter}, {Aldering}, {Goldhaber},
  {Knop}, {Nugent}, {Castro}, {Deustua}, {Fabbro}, {Goobar}, {Groom}, {Hook},
  {Kim}, {Kim}, {Lee}, {Nunes}, {Pain}, {Pennypacker}, {Quimby}, {Lidman},
  {Ellis}, {Irwin}, {McMahon}, {Ruiz-Lapuente}, {Walton}, {Schaefer}, {Boyle},
  {Filippenko}, {Matheson}, {Fruchter}, {Panagia}, {Newberg}, {Couch}, \&
  {Supernova Cosmology Project}}]{Perlmutter99}
{Perlmutter}, S., {Aldering}, G., {Goldhaber}, G., {Knop}, R.~A., {Nugent}, P.,
  {Castro}, P.~G., {Deustua}, S., {Fabbro}, S., {Goobar}, A., {Groom}, D.~E.,
  {Hook}, I.~M., {Kim}, A.~G., {Kim}, M.~Y., {Lee}, J.~C., {Nunes}, N.~J.,
  {Pain}, R., {Pennypacker}, C.~R., {Quimby}, R., {Lidman}, C., {Ellis}, R.~S.,
  {Irwin}, M., {McMahon}, R.~G., {Ruiz-Lapuente}, P., {Walton}, N., {Schaefer},
  B., {Boyle}, B.~J., {Filippenko}, A.~V., {Matheson}, T., {Fruchter}, A.~S.,
  {Panagia}, N., {Newberg}, H.~J.~M., {Couch}, W.~J., \& {Supernova Cosmology
  Project}. 1999, \apj, 517, 565

\bibitem[{{Riess} {et~al.}(1998){Riess}, {Filippenko}, {Challis},
  {Clocchiatti}, {Diercks}, {Garnavich}, {Gilliland}, {Hogan}, {Jha},
  {Kirshner}, {Leibundgut}, {Phillips}, {Reiss}, {Schmidt}, {Schommer},
  {Smith}, {Spyromilio}, {Stubbs}, {Suntzeff}, \& {Tonry}}]{Riess98}
{Riess}, A.~G., {Filippenko}, A.~V., {Challis}, P., {Clocchiatti}, A.,
  {Diercks}, A., {Garnavich}, P.~M., {Gilliland}, R.~L., {Hogan}, C.~J., {Jha},
  S., {Kirshner}, R.~P., {Leibundgut}, B., {Phillips}, M.~M., {Reiss}, D.,
  {Schmidt}, B.~P., {Schommer}, R.~A., {Smith}, R.~C., {Spyromilio}, J.,
  {Stubbs}, C., {Suntzeff}, N.~B., \& {Tonry}, J. 1998, \aj, 116, 1009

\bibitem[{{Schmidt} {et~al.}(1998){Schmidt}, {Suntzeff}, {Phillips},
  {Schommer}, {Clocchiatti}, {Kirshner}, {Garnavich}, {Challis}, {Leibundgut},
  {Spyromilio}, {Riess}, {Filippenko}, {Hamuy}, {Smith}, {Hogan}, {Stubbs},
  {Diercks}, {Reiss}, {Gilliland}, {Tonry}, {Maza}, {Dressler}, {Walsh}, \&
  {Ciardullo}}]{Schmidt98}
{Schmidt}, B.~P., {Suntzeff}, N.~B., {Phillips}, M.~M., {Schommer}, R.~A.,
  {Clocchiatti}, A., {Kirshner}, R.~P., {Garnavich}, P., {Challis}, P.,
  {Leibundgut}, B., {Spyromilio}, J., {Riess}, A.~G., {Filippenko}, A.~V.,
  {Hamuy}, M., {Smith}, R.~C., {Hogan}, C., {Stubbs}, C., {Diercks}, A.,
  {Reiss}, D., {Gilliland}, R., {Tonry}, J., {Maza}, J., {Dressler}, A.,
  {Walsh}, J., \& {Ciardullo}, R. 1998, \apj, 507, 46

\bibitem[{{Schwab} {et~al.}(2010){Schwab}, {Spronck}, {Tokovinin}, \&
  {Fischer}}]{Schwab10}
{Schwab}, C., {Spronck}, J.~F.~P., {Tokovinin}, A., \& {Fischer}, D.~A. 2010,
  in Society of Photo-Optical Instrumentation Engineers (SPIE) Conference
  Series, Vol. 7735, Society of Photo-Optical Instrumentation Engineers (SPIE)
  Conference Series

\bibitem[{{Seidelmann}(1992)}]{Seidelmann92}
{Seidelmann}, P.~K. 1992, {Explanatory Supplement to the Astronomical Almanac}
  (University Science Books)

\bibitem[{{Simms} {et~al.}(2005){Simms}, {Burke}, {Claver}, {Heathcote},
  {Rosenberg}, {Asztalos}, {Becker}, {Britton}, {Ellerbroek}, {Gilmore},
  {Hainaut-Rouelle}, {Jernigan}, {Kahn}, {Krabbendam}, {Margoniner}, {Monet},
  {Peterson}, {Pinto}, {Puxley}, {Rasmussen}, {Sebag}, {Tokovinin}, {Tyson}, \&
  {Wittman}}]{Simms05}
{Simms}, L., {Burke}, D., {Claver}, C.~F., {Heathcote}, S., {Rosenberg}, L.,
  {Asztalos}, S., {Becker}, A., {Britton}, M., {Ellerbroek}, B., {Gilmore}, K.,
  {Hainaut-Rouelle}, M.-C., {Jernigan}, G., {Kahn}, S.~M., {Krabbendam}, V.,
  {Margoniner}, V., {Monet}, D., {Peterson}, J., {Pinto}, P., {Puxley}, P.,
  {Rasmussen}, A., {Sebag}, J., {Tokovinin}, A., {Tyson}, J.~A., \& {Wittman},
  D. 2005, in Bulletin of the American Astronomical Society, Vol.~37, American
  Astronomical Society Meeting Abstracts, 1206

\bibitem[{{Smith} {et~al.}(1994){Smith}, {Kolenkiewicz}, {Nerem}, {Dunn},
  {Torrence}, {Robbins}, {Klosko}, {Williamson}, \& {Pavlis}}]{Smith94}
{Smith}, D.~E., {Kolenkiewicz}, R., {Nerem}, R.~S., {Dunn}, P.~J., {Torrence},
  M.~H., {Robbins}, J.~W., {Klosko}, S.~M., {Williamson}, R.~G., \& {Pavlis},
  E.~C. 1994, Geophysical Journal International, 119, 511

\bibitem[{{Tapley} {et~al.}(1985){Tapley}, {Schutz}, \& {Eanes}}]{Tapley85}
{Tapley}, B.~D., {Schutz}, B.~E., \& {Eanes}, R.~J. 1985, Celestial Mechanics,
  37, 247

\bibitem[{{U.~S.~Government Printing Office (USGPO)}(2013)}]{AA2013}
{U.~S.~Government Printing Office (USGPO)}. 2013, {The Astronomical Almanac for
  the year 2013}

\bibitem[{{Vilas} \& {Lasker}(1977)}]{Vilas77}
{Vilas}, F. \& {Lasker}, B.~M. 1977, \pasp, 89, 95

\bibitem[{{Walker}(1980)}]{Walker80}
{Walker}, A.~R. \& Mu\~{n}oz, J. 1980, {Facilities manual of the Cerro Tololo
  Inter-American Observatory (CTIO)}

\bibitem[{{Zombeck}(2007)}]{Zombeck07}
{Zombeck}, M. 2007, {Handbook of Space Astronomy and Astrophysics: Third
  Edition} (Cambridge University Press)

\end{thebibliography}

\bibliographystyle{apj}

\vspace*{1.0cm}
\noindent {\bf Updates}\\
4 Oct 2012: Posted to arXiv: \href{http://xxx.lanl.gov/abs/1210.1616}{http://xxx.lanl.gov/abs/1210.1616}\\
9 Oct 2012: Minor edits. Added hyperlinks for all websites. Added Google Maps hyperlinks for all coordinates listed in 
Tables 1, 4, and 6. Table 6: Added table notes. Added footnote on LSST floor elevation. No numbers were changed.\\
1 Mar 2013: Slight revisions for elevations for Gemini and SOAR in Table 6, with explanations added in table notes $e$ and $f$. Minor revisions to text in Introduction.\\

\newpage

\begin{deluxetable*}{lllllllclcllll}
\tabletypesize{\scriptsize}
\tablecaption{Geodetic and Geocentric Coordinates for Observatories on\\Cerro Tololo and Cerro Pach\'{o}n\label{tab:final}}
\tablewidth{0pt}
\tablehead{
\colhead{Site} & \multicolumn{3}{c}{~~$\phi_{geodetic}$~~} & \multicolumn{3}{c}{~~$\lambda_{geodetic}$~~} & \colhead{~Ref.~} & \colhead{~~$h$~~} & \colhead{~~~$N$~~~} & \colhead{~~$H$~~} & \multicolumn{3}{c}{~~$\phi^{'}_{geocentric}$~~} \\ 
\colhead{} & \colhead{$\circ$} & \colhead{'} & \colhead{''} & \colhead{$\circ$} & \colhead{'} & \colhead{''} & \colhead{} & \colhead{(m)} & \colhead{(m)} & \colhead{(m)} & \colhead{$\circ$} & \colhead{'} & \colhead{''} }
\startdata
{\bf Cerro Tololo}     &   &  &     &   &  &     &   &     &       &   &  &  \\
\hline                                                
\href{https://maps.google.com/maps?&z=18&q=-30.1696611,-70.8065250}{Blanco 4-m} &-30&10&10.78&-70&48&23.49& 1 &2241.4$^a$ &34.6&2206.8$^a$ &-30&00&10.03\\
\href{https://maps.google.com/maps?&z=18&q=-30.1692833,-70.8067889}{SMARTS 1.5-m} &-30&10&09.42&-70&48&24.44& 1 &2241.9$^{a,b}$ &34.6&2207.3$^{a,b}$ &-30&00&08.68\\
\href{https://maps.google.com/maps?&z=18&q=-30.1688667,-70.8060639}{SMARTS 1.0-m} &-30&10&07.92&-70&48&21.83& 1 &2240.5$^a$ &34.6&2205.9$^a$ &-30&00&07.18\\
\href{https://maps.google.com/maps?&z=18&q=-30.1688611,-70.8066277}{SMARTS 0.9-m} &-30&10&07.90&-70&48&23.86& 1 &2241.4$^a$ &34.6&2206.8$^a$ &-30&00&07.16\\
\href{https://maps.google.com/maps?&z=18&q=-30.1690556,-70.8062861}{Curtis Schmidt 0.6-m} &-30&10&08.60&-70&48&22.63& 1 &2240.9$^a$ &34.6&2206.3$^a$ &-30&00&07.86\\
\href{https://maps.google.com/maps?&z=18&q=-30.1685972,-70.8063389}{Former UCAC} &-30&10&06.95&-70&48&22.82& 1 &2241.2$^a$ &34.6&2206.6$^a$ &-30&00&06.21\\
\href{https://maps.google.com/maps?&z=18&q=-30.1686028,-70.8060139}{Former CHASE} &-30&10&06.97&-70&48&21.65& 1 &2240.5$^a$ &34.6&2205.9$^a$ &-30&00&06.23\\
\href{https://maps.google.com/maps?&z=18&q=-30.1688972,-70.8069750}{RASICAM}    &-30&10&08.03&-70&48&25.11& 1 &2238$^a$ &34.6&2204$^a$ &-30&00&07.29\\
\href{https://maps.google.com/maps?&z=18&q=-30.1683111,-70.8035694}{WHAM}       &-30&10&05.92&-70&48&12.85& 1 &2188$^{c,d}$ &34.7&2153$^{c,d}$ &-30&00&05.18\\
\href{https://maps.google.com/maps?&z=18&q=-30.1671778,-70.8039972}{KASI}       &-30&10&01.84&-70&48&14.39& 2 &2182$^{c,d}$ &34.7&2147$^{c,d}$ &-30&00&01.12\\
\href{https://maps.google.com/maps?&z=18&q=-30.1673833,-70.8047889}{LCOGT Stellan-A} &-30&10&02.58&-70&48&17.24& 1 &2198$^{c,d}$ &34.6&2163$^{c,d}$ &-30&00&01.85\\
\href{https://maps.google.com/maps?&z=18&q=-30.1673306,-70.8046611}{LCOGT Stellan-B} &-30&10&02.39&-70&48&16.78& 1 &2198$^{c,d}$ &34.6&2163$^{c,d}$ &-30&00&01.66\\
\href{https://maps.google.com/maps?&z=18&q=-30.1674472,-70.8046806}{LCOGT Stellan-C} &-30&10&02.81&-70&48&16.85& 1 &2198$^{c,d}$ &34.6&2163$^{c,d}$ &-30&00&02.08\\
\href{https://maps.google.com/maps?&z=18&q=-30.1674472,-70.8049778}{SMARTS 1.3-m} &-30&10&02.81&-70&48&17.92& 1 &2200$^{c,d}$ &34.6&2165$^{c,d}$ &-30&00&02.08\\
\href{https://maps.google.com/maps?&z=18&q=-30.1676444,-70.805252777}{PROMPT \#1} &-30&10&03.52&-70&48&18.91& 3 &2207$^{c,d}$ &34.6&2172$^{c,d}$ &-30&00&02.79\\
\href{https://maps.google.com/maps?&z=18&q=-30.1676361,-70.805433333}{PROMPT \#2} &-30&10&03.49&-70&48&19.56& 3 &2207$^{c,d}$ &34.6&2172$^{c,d}$ &-30&00&02.76\\
\href{https://maps.google.com/maps?&z=18&q=-30.16754722,-70.80523611}{PROMPT \#3} &-30&10&03.17&-70&48&18.85& 3 &2207$^{c,d}$ &34.6&2172$^{c,d}$ &-30&00&02.44\\
\href{https://maps.google.com/maps?&z=18&q=-30.16765556,-70.80536667}{PROMPT \#4} &-30&10&03.56&-70&48&19.32& 1 &2207$^{c,d}$ &34.6&2172$^{c,d}$ &-30&00&02.83\\
\href{https://maps.google.com/maps?&z=18&q=-30.16754444,-70.80532222}{PROMPT \#5} &-30&10&03.16&-70&48&19.16& 3 &2207$^{c,d}$ &34.6&2172$^{c,d}$ &-30&00&02.43\\
\href{https://maps.google.com/maps?&z=18&q=-30.16772778,-70.80526667}{PROMPT \#6} &-30&10&03.82&-70&48&18.96& 3 &2207$^{c,d}$ &34.6&2172$^{c,d}$ &-30&00&03.09\\
\href{https://maps.google.com/maps?&z=18&q=-30.16786944,-70.80537778}{PROMPT \#7} &-30&10&04.33&-70&48&19.36& 1 &2208$^{c,d}$ &34.6&2173$^{c,d}$ &-30&00&03.60\\
\href{https://maps.google.com/maps?&z=18&q=-30.16775556,-70.80551111}{GONG} &-30&10&03.92&-70&48&19.84& 1 &2209$^{c,d}$ &34.6&2174$^{c,d}$ &-30&00&03.19\\
\href{https://maps.google.com/maps?&z=18&q=-30.16813333,-70.80543333}{SSI Airglow} &-30&10&05.28&-70&48&19.56& 1 &2212$^{c,d}$ &34.6&2177$^{c,d}$ &-30&00&04.55\\
\href{https://maps.google.com/maps?&z=18&q=-30.16786389,-70.80568889}{T80-South (site)} &-30&10&04.31&-70&48&20.48& 2 &2212$^{c,d}$ &34.6&2178$^{c,d}$ &-30&00&03.58\\
\href{https://maps.google.com/maps?&z=18&q=-30.17214444,-70.799202778}{SARA South 0.6-m} &-30&10&19.72&-70&47&57.13& 1 &2151$^{c,d}$ &34.7&2116$^{c,d}$ &-30&00&18.93\\
\hline                                                
{\bf Cerro Pach\'{o}n}   & & &   & & &   & &     &   &     & & & \\
\hline                                                
\href{https://maps.google.com/maps?&z=18&q=-30.2407416667,-70.7366833}{Gemini South 8.2-m} &-30&14&26.67&-70&44&12.06& 1 &2748$^e$ &35.1&2713$^e$ &-30&04&25.12\\
\href{https://maps.google.com/maps?&z=18&q=-30.2378916667,-70.73364167}{SOAR 4.1-m} &-30&14&16.41&-70&44&01.11& 1 &2738$^f$ &35.1&2703$^f$ &-30&04&14.89\\
\href{https://maps.google.com/maps?&z=18&q=-30.2446333333,-70.749416667}{LSST 8.4-m (site)} &-30&14&40.68&-70&44&57.90& 2 &2647$^g$ &35.0&2612$^g$ &-30&04&39.07\\
\href{https://maps.google.com/maps?&z=18&q=-30.2447972222,-70.747722222}{LSST Aux. 1.4-m (site)} &-30&14&41.27&-70&44&51.80& 2 &2647$^g$ &35.0&2612$^g$ &-30&04&39.66\\
\href{https://maps.google.com/maps?&z=18&q=-30.2517694444,-70.738194444}{Andes LIDAR Obs. (ALO)} &-30&15&06.37&-70&44&17.50& 1 &2552$^{c,d}$ &35.1&2517$^{c,d}$ &-30&05&04.67
\enddata

\tablecomments{Coordinate reference codes: (1) average of GPS and
  corrected Google~Earth positions, (2) GPS position, (3) corrected
  Google~Earth position.  Geodetic latitude $\phi$, longitude
  $\lambda$ on WGS~84 system.  Geodetic elevation $h$ is with respect
  to the WGS~84 ellipsoid.  Elevation $H$ is orthometric height with
  respect to geoid (i.e.  above ``mean sea level''; $H$ = $h$ -
  $N$). $N$ is geoid undulation according to EGM-96 model, i.e. it is
  the height of the geoid above the WGS~84 reference ellipsoid.
  Geocentric latitude $\phi^{'}$ was calculated following \S4 of
  \citet{Seidelmann92}.  Total uncertainties in latitude and longitude
  are approximately $\pm$0''.10.  Elevations are ground level at the
  site, {\it not} the elevation of the telescope.}

\tablenotetext{a}{Ground elevations of facilities on Tololo plateau
  are likely accurate to better than $\pm$1 m absolute (calculated
  using 2008 survey geodetic height for Tololo Control survey monument
  and differential heights from 1973 survey).}

\tablenotetext{b}{SMARTS 1.5-m: According to building plans for the
  1.5-m, the center of its dome is 10.3\,m above ground level. So the
  1.5-m telescope is at $h$ $\simeq$ 2252\,m and elevation $H$
  $\simeq$ 2218\,m.}

\tablenotetext{c}{Elevations of other Tololo facilities and ALO are a
  blend of corrected GPS and Google Earth elevations, but should be
  better than $\pm$5 m (rms).}

\tablenotetext{d}{The elevations listed for facilities on the slopes
  of Cerro Tololo are commensurate with that of a
  \href{https://maps.google.com/maps?&z=18&q=-30.16778194444,-70.80510155556}{survey
    monument labeled ``{\it Pilote 1}''} ($h_{GRS~80}$ = 2205.415 m
  $\simeq$ $h_{WGS~84}$) situated in elevation between the SMARTS
  1.3-m and PROMPT cluster.}

\tablenotetext{e}{Gemini South: The elevation is adopted from the 2008
  survey value for the survey marker ``Pachon IGM'' ($h$ = 2648 m)
  which is in the vicinity of the observatory structure. The marker
  elevation agrees well with the author's {\it corrected} GPS
  elevation ($h$ = 2650 m) and {\it corrected} Google Earth elevation
  ($h$ = 2646 m). Systematic error in $h$ and $H$ for Gemini South are
  unlikely to exceed $\pm$2 m.}

\tablenotetext{f}{SOAR: The elevation was adopted from the SOAR
  website
  \href{http://www.soartelescope.org/observing/visiting-astronomers-guide}{http://www.soartelescope.org/observing/visiting-astronomers-guide}($h$
  = 2738 m), which agrees well with author's {\it corrected} GPS value
  ($h$ = 2740 m). The {\it corrected} Google Earth elevation ($h$ =
  2723 m) is somewhat lower than the other two estimates, however this
  may be due to the extreme narrowness of the peak upon which SOAR is
  situated, due to the low resolution of the Google Earth elevation
  maps.}

\tablenotetext{g}{LSST and LSST Auxiliary Telescope: elevations are
  adopted from LSST website
  \href{http://www.lsst.org/lsst/science/summit_facilities}{http://www.lsst.org/lsst/science/summit$\_$facilities},
  which were based on cartography tied to the 2008 survey. According
  to current LSST plans, the pier floor will be at $h$ = 2662.75\,m.}

\end{deluxetable*}

\end{document}